\documentclass[fleqn,usenatbib]{mnras}

\usepackage{newtxtext,newtxmath}
\usepackage{textcomp}

\usepackage[T1]{fontenc}

\usepackage{graphicx}	
\usepackage{amsmath}	
\usepackage{physics}

\title[]{Modelling magnetically formed neutron star mountains}

\author[A. Nanda]{
Amlan Nanda$^{1,2}$\thanks{E-mail: amlan.nanda@wsu.edu}
\\
$^{1}$Research Center for the Early Universe (RESCEU), School of Science, The University of Tokyo, 7-3-1 Hongo, Bunkyo, Tokyo 113-0033, Japan \\
$^{2}$Department of Physics and Astronomy, Washington State University, Pullman, Washington 99164–2814, USA
}
\date{Accepted XXX. Received YYY; in original form ZZZ}

\pubyear{2024}

\begin{document}
\label{firstpage}
\pagerange{\pageref{firstpage}--\pageref{lastpage}}
\maketitle

\begin{abstract}
With the onset of the era of gravitational-wave (GW) astronomy, the search for continuous gravitational waves (CGWs), which remain undetected to date, has intensified in more ways than one. Rapidly rotating neutron stars with non-axisymmetrical deformations are the main targets for CGW searches. The extent of this quadrupolar deformation is measured by the maximum ellipticity that can be sustained by the crust of a neutron star and it places an upper limit on the CGW amplitudes emitted by such systems. In this paper, following previous works on this subject, we calculate the maximum ellipticity of a neutron star generated by the Lorentz force exerted on it by the internal magnetic fields. We show that the ellipticity of stars deformed by such a Lorentz force is of the same order of magnitude as previous theoretical and astrophysical constraints. We also consider if this ellipticity can be further enhanced by crustal surface currents. We discover that this is indeed true; surface currents at crustal boundaries are instrumental towards enhancing the ellipticity of magnetized neutron stars. 
\end{abstract}



\section{Introduction}

Neutron stars have been popular targets in Gravitational-wave (GW) astronomy  (\citealt{10.1093/mnras/184.3.501}; \citealt{1984ApJ...278..345W}) due to their extreme compactness and multiple energetic events in the universe associated with them. They can radiate GWs through multiple different astrophysical phenomena; including in the form of binary inspiral and merger (\citealt{PhysRevD.85.022001}), through various modes of oscillation and instabilities (\citealt{1992PhRvD..46.4289K}; \citealt{1998ApJ...502..708A}; \citealt{andersson1999relevance}), and as rotating neutron stars deformed away from axisymmetry (\citealp{1998ApJ...501L..89B}). The first detection of a binary black hole merger (\citealp{Abbott_2016}) by the GW detectors of the LIGO Scientific Collaboration heralded the beginning of the era of GW astronomy, opening up an entirely new channel to explore the universe. The first detection of a binary neutron star merger (\citealp{PhysRevLett.119.161101}) along with its electromagnetic counterparts (\citealp{Abbott_2017}) generated reinvigorated interest in neutron stars as sources of GWs, driving renewed efforts in exploring neutron stars as sources of GWs.

The next highly anticipated discovery is continuous gravitational waves (CGWs) from neutron stars with non-axisymmetric deformations (\citealt{Sieniawska_2019}; \citealt{wette2023searches}). The motivation behind expecting this observation stems from the observation (\citealt{doi:10.1126/science.1123430}; \citealt{2007PhR...442..109L}) that no neutron star spins close to the centrifugal break-up frequency which is generally above $\sim$ 1 kHz for most Equation-of-State (EoS) candidates (\citealt{1994ApJ...424..823C}). It was suggested that the lack of neutron stars rotating at these extreme rates is due to the emission of CGWs, which provide a braking torque that halts the spin-up of neutron stars (\citealt{1998ApJ...501L..89B}; \citealt{andersson1999relevance}; \citealt{Gittins_2019}). The associated non-axisymmetric (quadrupolar) deformations of neutron stars are commonly referred to as \emph{mountains}. 

CGW detection strategies from rapidly rotating neutron stars have been three-pronged: 1) targeted observation of specific pulsars for signatures of CGWs (\citealt{abbott2019narrow}; \citealt{Piccinni_2020}; \citealt{Ashok_2021}; \citealt{abbott2021diving}; \citealt{abbott2022search}; \citealt{Abbott_2022}; \citealt{abbott2022narrowband}; \citealt{abbott2022searches}); 2) all-sky wide-parameter surveys for unobserved sources (\citealt{abbott2017first}; \citealt{abbott2021all}; \citealt{2021A&A...649A..92C}; \citealt{Dergachev_2021}; \citealt{abbott2022all}; \citealt{steltner2023deep}); and 3) CGW observations using directed searches (\citealt{aasi2015searches}; \citealt{Zhu_2016}; \citealt{Lindblom_2020}; \citealt{Papa_2020}; \citealt{abbott2021searches}, \citealt{PhysRevD.106.062002}, \citealt{PhysRevD.106.042003}, \citealt{LIGOScientific:2022enz}). A measure of the size of these mountains is the ellipticity of the neutron star \citet{1998ApJ...501L..89B} and ellipticities of multiple observed pulsars have been constrained to less than $10^{-8}$ (\citealt{abbott2020gravitational}). It is thus important to estimate the size of the largest mountain that a neutron star crust can support, which would also provide an upper limit on the amplitude of CGWs emitted from such systems. The size of this mountain gives us an estimate of the expected amplitude of the CGWs such a star would emit (\citealt{Ushomirsky_2000}) and hence, this can help in constructing detection templates for CGWs targeted by the runs of the LIGO-Virgo-KAGRA collaboration (\citealt{kiendrebeogo2023updated}; \citealt{miller2023recent}), which can then be supplemented through electromagnetic observations of known pulsars (\citealt{falxa2023searching}).
 
The earliest known attempt at calculating the maximum quadrupolar deformation of a neutron star in Newtonian gravity was conducted by \cite{Ushomirsky_et_al_2000}, who utilized the Cowling approximation to derive an expression for the quadrupole moment, assuming that the star will sustain its maximum mountain when the crust is maximally strained to the elastic yield limit at every point. Their approach had an issue, however; their strain tensor calculation did not account for the boundary conditions at the crustal interfaces (\citealt{Haskell_2006}). This calculation was then improved by \cite{Haskell_2006}, which was again corrected recently by \cite{2021MNRAS.500.5570G}, where they calculated the maximum mountain sustained by the crust of a neutron star when subjected to a deforming force, which they considered to be of three distinct types. This approach also accounted for satisfying the boundary conditions of this particular force, especially at the crustal interfaces of the star. \cite{Morales_2022} built on this work to construct a generalized force with the aim of maximizing the crustal mountain of a neutron star with the aim of reconciling the conclusions of \cite{Ushomirsky_et_al_2000} and \cite{2021MNRAS.500.5570G} regarding their estimates of the ellipticities of neutron stars. A general relativistic extension of this exercise was provided in \cite{Gittins_2021}, which demonstrated ellipticity estimates were further suppressed compared to its Newtonian counterpart. 

In this paper, we calculate the size of the \emph{magnetic mountain} that can be sustained on the crust of a neutron star. These mountains are assumed to be generated solely by the internal magnetic fields of the neutron star (\citealt{article, Mastrano_2015}), which are assumed to be extremely strong with a magnitude of $\sim 10^{15} \text{G}$ (\citealt{1992ApJ...392L...9D}; \citealt{Kaspi_2017}). 
Contrary to \cite{2021MNRAS.500.5570G} and \citet{Morales_2022}, who use a proxy force for all the internal dynamics of the star during its evolutionary history, we use the Lorentz force generated by the internal magnetic fields of the neutron star to estimate the maximum size of the \emph{magnetic mountains}. Similar to \cite{2021MNRAS.500.5570G} and \citet{Morales_2022}, our calculations are also done completely in Newtonian gravity with the aim of providing a reasonable order of magnitude estimate for the size of these mountains. We also calculate the size of \emph{magnetic mountains} generated due to internal magnetic fields of the star in the presence of surface currents(\citealt{10.1093/mnras/stt1008}; \citealt{viganò2015magnetic}; \citealt{Anzuini_2022}; \citealt{Stefanou_2022}).
We demonstrate that the surface currents of the neutron stars present at the crustal boundaries of the star will generate highly deformed mountains that can be sustained on the crust of the neutron star. This is motivated by the result that deforming forces (in our case, they are the Lorentz forces generated by the modified magnetic fields) which are of increased magnitude in the crust of the star compared to the core and the ocean can induce higher ellipticities in the star (\citealt{Morales_2022}). 

We model the internal structure of our neutron star using a simple three-layered structure; we assume it has a fluid core which makes up the majority of the volume of the star, on top of which we have the elastic crust of the star that is about $\sim 1 \text{km}$ thick, which has a very thin fluid ocean on top of it (\citealt{bildsten1997oceanography}) which marks the surface of the star. 
After the neutron star is born (\citealt{1973NPhS..246...93R}; \citealt{Lander_2021}; \citealt{Varma_2022}; \citealt{White_2022}), in its infancy the star is completely fluid after which its crust solidifies in a characteristic timescale (\citealt{2014PASJ...66L...1S}; \citealt{2020ApJ...888...97B}). It is at this stage the extremely strong internal magnetic fields of the star deform it before they decay to lower values. 
To explain how mountains are formed in detail, we rely on the mountain building scheme developed by \cite{2021MNRAS.500.5570G}. We build a system of coupled linear differential equations that describe the perturbation of the parameters describing the fluid core, fluid ocean and elastic crust of the star due to the Lorentz force generated by the internal magnetic fields of the star in the first case, and the modified Lorentz force when the surface currents are also present at the crustal boundaries. We supplement these with the proper boundary conditions and solve this system of perturbative equations numerically. 

This paper is organized as follows. In Section \ref{section:2}, we describe our background star and discuss the perturbation framework we use in this paper. In Section \ref{section:3}, we discuss the mountain building scheme that we use for the two different scenarios that we consider in this paper, focusing on both the mathematical formulation required as well as an intuitive understanding of the problem. In Section \ref{section:4}, we describe the equations governing the internal magnetic fields and the crustal surface currents of our star for the 2 aforementioned scenarios. We then begin the exercise of perturbing our star and calculating its ellipticity and the extent of its quadrupolar deformation in section \ref{section:5}. Finally, we solve each of these systems of perturbative equations numerically in \ref{section:6} to estimate the maximum possible ellipticity of the star in both scenarios. Finally, we discuss our results, suggest future possible work and state the importance of this work in Section \ref{section:7}.

\section{Background Star and Perturbation framework}\label{section:2}
\subsection{The Background star}
We first describe the equations governing our background, an unperturbed fluid neutron star. Following the example of \cite{2021MNRAS.500.5570G}, we restrict ourselves to Newtonian gravity, enabling us to assume a simple polytropic equation of state (EoS) for the background star. We assume our background star is a non-rotating, barotropic fluid star with density $\rho$, isotropic pressure $p$, and gravitational potential $\Phi$. 
Since the star is in equilibrium and our system is taken to be static, 
a barotropic fluid is governed by the following equations: 
\begin{equation}\label{eq:1}
    \frac{\partial\rho}{\partial t} + \Vec{\nabla}\cdot(\rho\Vec{v}) = 0,
\end{equation}
\begin{equation}\label{eq:3}
    \Vec{\nabla}p + \rho\Vec{\nabla}\Phi = 0,
\end{equation} 
where the gravitational potential is defined using Poisson's equation 
\begin{equation}
    \nabla^{2}\Phi = 4\pi G\rho.
\end{equation} 
We use a polytropic Equation of State of the form 
\begin{equation}\label{EoSp}
    p(\rho) = K\rho^{1+\frac{1}{n}},
\end{equation} where $K$ is a polytropic constant of proportionality and $n$ is the polytropic index. For all our intents and purposes, we use the $n=1$ polytropic EoS model for the background star, and the density profile becomes
\begin{equation}\label{EoS}
    \rho(r) = \rho_c\frac{\sin(\pi r/R)}{\pi r/R},
\end{equation}
where the $\rho_c$ and $R$ are the central density and the stellar radius. The stellar mass and radius are determined by the central density and the polytropic constant $K$.

We use CGS units throughout all calculations. We assumed the central density of the star and the polytropic constant $K$ is $\rho_{c} = 2.185 \times 10^{15} \text{g}\text{cm}^{-3}$ and $K = 42500$. The radius of the star is $R = 1.00011 \times 10^{6} \text{cm}$. The value of the universal gravitational constant in CGS units is $G = 6.6743 \cross 10^{-8} \text{cm}^{3}g^{-1}s^{-2}$. The mass of the star is estimated numerically from the background equation of the neutron star \begin{equation}
    \frac{dm}{dr} = 4\pi r^{2}\rho, 
\end{equation} 
where $m$ is the enclosed mass.
We thus obtain the mass of the star $M_{NS} = 1.4 \text{M}_\odot$. Now, we calculate the location of the crustal boundaries in our neutron star, i.e the demarcation between the core and the crust of the star ($r = r_{1}$) and the demarcation between the crust and the ocean of the star ($r = r_{2}$). We implement this calculation numerically by using an iterative procedure where we try to match the value of densities at these demarcations to their corresponding radii, which is essentially inverting equation (\ref{EoS}). We assume the density of the star at the core-crust boundary to be $\rho_{cc} = 2.0 \cross 10^{14} \text{g}/\text{cm}^{3}$ and thus obtain the core-crust boundary to be at $r_{1} = 0.91533 \times 10^{6} \text{cm}$. Similarly, we assume the density of the star at the crust-ocean boundary to be $\rho_{co} = 1.0 \times 10^{11} \text{g}/\text{cm}^{3}$ and thus obtain the crust-ocean boundary to be at $r_{2} = 1.00007 \times 10^{6} \text{cm}$. This gives us the complete description of the background star that is needed for our calculations.
\subsection{Perturbation framework}

Now, when there is a force perturbing these parameters of the star, we model the perturbed parameters using a Lagrangian perturbation framework (\citealt{1978ApJ...221..937F}) by introducing the Lagrangian displacement vector $\Vec{\xi}$. In a static background star, it is expressed as \begin{equation}
    \frac{\partial\Vec{\xi}}{\partial t} = \Delta\Vec{v} = \delta\Vec{v}.
\end{equation} 
Our perturbation equations are thus written as 
\begin{equation}\label{eq:delrho}
    \delta\rho = -\Vec{\nabla}\cdot(\rho\Vec{\xi}),
\end{equation}\begin{equation}\label{eq:2}
    \Vec{\nabla}\delta p + \delta\rho\Vec{\nabla}\Phi + \rho\Vec{\nabla}\delta\Phi = 0,
\end{equation}\begin{equation}\label{eq:delp}
    \delta p = c^{2}_{s}\delta\rho,
\end{equation}\begin{equation}\label{eq:8}
    \nabla^{2}\delta\Phi = 4\pi G\delta\rho,
\end{equation} where $c^{2}_{s} = \frac{\partial p}{\partial\rho}$ is the square of the speed of sound in the star. Since we focus on perturbations of the star induced by the Lorentz force, we modify the perturbed Euler equation (equation (\ref{eq:2})) to \begin{equation}\label{eq:7}
    \Vec{\nabla}\delta p + \delta\rho\Vec{\nabla}\Phi + \rho\Vec{\nabla}\delta\Phi = \Vec{f}_{L},
\end{equation} where $\Vec{f}_{L}$ is the Lorentz force. We use the Lorentz force as an example of a force which satisfies all the boundary conditions of the problem as well as enables us to perturbatively model stars deformed away from axisymmetry. This is a key difference in our approach from \cite{2021MNRAS.500.5570G} and \citet{Morales_2022}; while we specifically consider the strong Lorentz force present in newly born neutron stars, in their case the force used was a proxy for the complicated processes occurring in the evolutionary history of a neutron star as it deforms away from axisymmetry. When we include the elastic crust in our star, we must account for the shear stresses inside the crust, and we, thus, modify our Euler equation (equation (\ref{eq:2})) again to obtain 
\begin{equation}\label{eq:5}
    \Vec{\nabla}\delta p + \delta\rho\Vec{\nabla}\Phi + \rho\Vec{\nabla}\delta\Phi - \Vec{\nabla}\cdot(\mathbf{t}) = \Vec{f}_{L},
\end{equation} 
where $\mathbf{t}$ is the shear-stress tensor (\citealt{Ushomirsky_2000}) which is assumed to enter the equations only in the perturbative order. This shear-stress tensor, characterizing the elastic crust, is written as \begin{equation}\label{eq:12}
    t_{ij} = \mu(\nabla_{i}\xi_{j} + \nabla_{j}\xi_{i} - \frac{2}{3}g_{ij}\nabla_{k}\xi^{k}),
\end{equation} where $\mu$ is the shear modulus of the crust and $g_{ij}$ is the three-metric in flat space. We consider a shear modulus profile of the following form in the crust (\citealt{Haskell_2006}) \begin{equation}\label{eq:kappa}
    \mu \equiv \mu(\rho) = \kappa\rho,
\end{equation} where $\kappa$ is a constant.

\section{Building Mountains}\label{section:3}
A star deformed away from a perfectly spherical shape develops multipole moments defined as 
\begin{equation}\label{quadrupoles}
    I_{lm} = \int^{R}_{0}\delta\rho_{lm}(r)r^{l+2}dr,
\end{equation}
where $(l,m)$ denotes the harmonics of the density perturbation $\delta\rho_{lm}$ and $R$ is the radius of the star. Note that we define our multipole moments using the perturbed density since we consider only perturbations of the star away from spherical geometry. We still need the background star to be spherical so that we can utilize the spherical harmonics (or the vector spherical harmonics (VSH)) basis to separate the radial and the angular variables in the problem. The full perturbative response is described as 
\begin{equation}
    \delta\rho (r,\theta,\phi) = \sum^{\infty}_{l=0}\sum^{l}_{m=-l}\delta\rho_{lm}(r)Y_{lm}(\theta,\phi),
\end{equation} 
as a sum over all the harmonic modes which contribute to the perturbation. Since we are interested in the CGW emission from the neutron star, the dominant multipole contributing to this emission is the quadrupole moment $I_{22}$. We focus solely on the quadrupolar harmonics denoted by $(l,m) = (2,2)$ for all our perturbations. What this essentially means is that for any internal parameter $Q$ of the star expanded using the spherical harmonics basis 
\begin{equation}
    Q(r,\theta,\phi) = \sum^{\infty}_{l=0}\sum^{l}_{m=-l}Q_{lm}(r)Y_{lm}(\theta,\phi),    
\end{equation}
we compute this expression's inner product with the $Y_{22}$ harmonic. We then obtain a system of perturbative equations with only the $Q_{22}$ parameters along with the necessary $l$-harmonic couplings. Another helpful parameter that we will also use in the following calculations is the fiducial ellipticity (\citealt{PhysRevLett.95.211101}) \begin{equation}\label{ellip}
    \epsilon = \sqrt{\frac{8\pi}{15}}\frac{I_{22}}{I_{zz}},
\end{equation} where $I_{zz}$ is the principal moment of inertia of the neutron star. 

Let us apply two different mountain-building schemes to the scenario that we are interested in; which is crustal deformations of a neutron star induced by magnetic fields. We modify our parameter set for each star state to include the magnetic fields. However, it doesn't enter the parameter set for the background star $\rho, p, \Phi$, but instead joins the perturbed parameter set ($\delta\rho, \delta p, \delta\Phi$), since we assume that all the other parameters are perturbed by the Lorentz force of internal origin in this scenario.

\begin{figure}
    \centering
    \includegraphics[width=9cm]{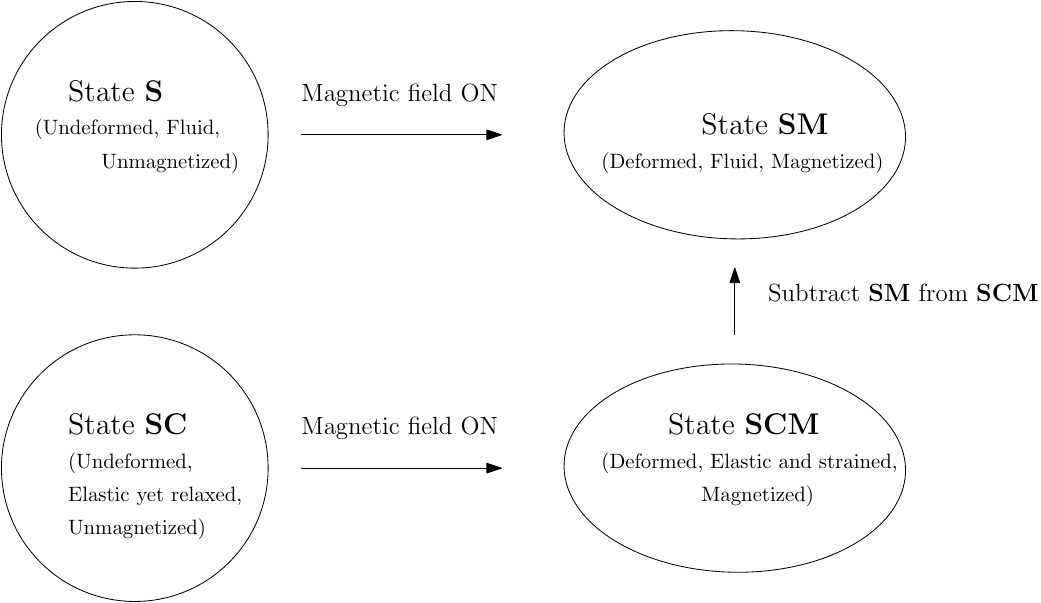}
    \caption{Formation of a magnetic mountain using the force-based approach.}
    \label{fig:1}
\end{figure}

We start with the scheme of mountain formation first motivated by \citet{2021MNRAS.500.5570G}. Suppose our background star state is that of an unmagnetized, and thereby undeformed, neutron star with a molten crust, which we call state \textbf{S}. Now we switch on the internal magnetic field of the star, 
and the fluid star is then deformed to a state labelled as \textbf{SM}. We take another star state, still unmagnetized and undeformed, but with its crust solidified and label it as state \textbf{SC}. We switch on the internal magnetic fields of extraordinary strength (\citealt{1992ApJ...392L...9D}) and this star is deformed into a star state labelled as \textbf{SCM}. We subtract star state \textbf{SM} from the state \textbf{SCM} which, based on our force-based approach to building mountains, gives us the perturbative equations (as detailed in section \ref{section:5}) to be solved to obtain the quadrupolar deformation of the neutron star. This scheme is illustrated by Fig. \ref{fig:1}.  

\begin{figure}
    \centering
    \includegraphics[width=9cm]{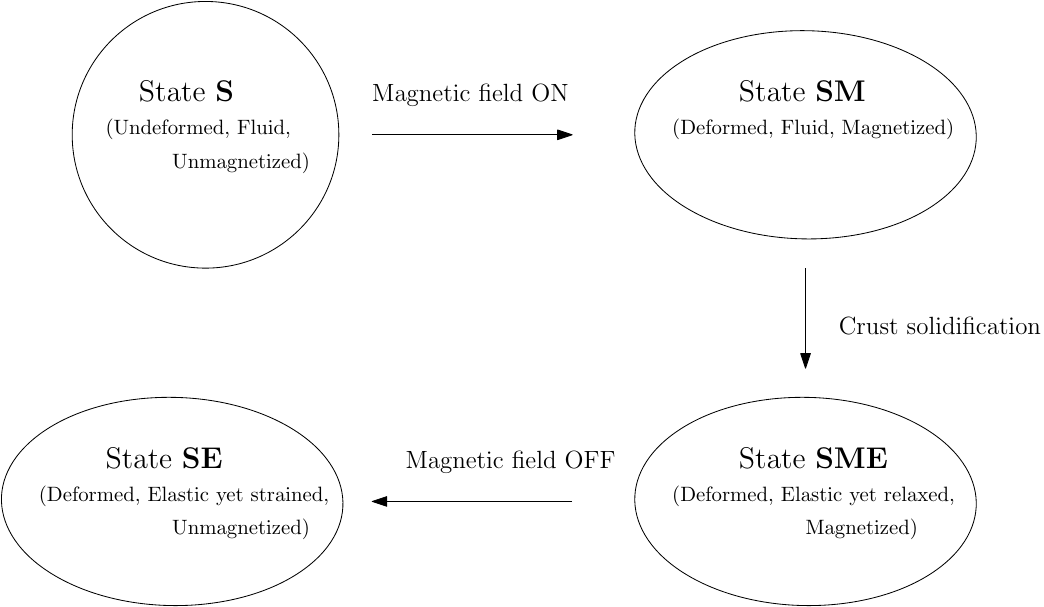}
    \caption{Formation of a magnetic mountain using the formation history-based approach.}
    \label{fig:2}
\end{figure}

Let's discuss how such mountains can form on neutron stars based on the formation history-based approach. This is necessary to physically motivate the formation of a magnetic mountain after the birth of a neutron star and to justify that there might be a reasonable population of neutron stars with such a mountain formed on their crust. When a young neutron star is formed, it is still undeformed (in a state labelled \textbf{S}) but with strong internal magnetic fields (\citealt{1992ApJ...392L...9D}). These magnetic fields exert Lorentz force on the star and deform this star, with its crust still molten, to a state \textbf{SM}. The crust of the neutron star solidifies in a timescale of a few tens of seconds to a day (\citealt{2014PASJ...66L...1S,2020ApJ...888...97B}) and the star is now still deformed but with an elastic crust, in a state labelled \textbf{SME}. We assume that the internal magnetic fields of a newly born neutron star decay 
on a much longer timescale (\citealt{Goldreich_Reisenegger_1992}), and the force deforming the neutron star has been removed and the star is in a state \textbf{SE}, with the crust strained to form a magnetic mountain. This scheme is illustrated by Fig. \ref{fig:2}. 

It is important to note that once the magnetic field deforms the molten crust from state \textbf{S} to \textbf{SM}, both of which are equilibrium states, the magnetic field is in MHD equilibrium with the fluid star and even after the crust of the neutron star solidifies in state \textbf{SME}, it remains in MHD equilibrium since the crust is still relaxed. This means that for both scenarios the solidification of the crust to star state \textbf{SME} doesn't disturb the MHD equilibrium of the magnetic fields with the fluid star. 

\section{Magnetic field configuration}\label{section:4}

In this section, we describe the internal magnetic field configuration for a neutron star based on various assumptions. We will see that in all cases the only factor determining this configuration is the equation of state of the star, which we have assumed to be an $n=1$ polytropic equation of state. 
We take our magnetic field to be of the form \begin{equation}\label{eq:4}
    \Vec{B} = B_{r}Y_{lm}\hat{r} + \frac{r}{\beta}B_{\perp}\Vec{\nabla}Y_{lm} + \frac{r}{\beta}B_{\cross}(\hat{r} \times \Vec{\nabla}Y_{lm}),
\end{equation} 
where $ \beta \equiv \sqrt{l(l+1)}$. This expression is a multipolar decomposition of the magnetic field using the vector spherical harmonics basis (\citealt{Carrascal_1991}). This decomposition is employed solely for the ease of perturbation calculations in the later sections. 
It is also worth noting that this decomposition is equivalent to the decomposition of stellar magnetic fields into poloidal and toroidal components (see Appendix \ref{appendix B}), which is popular in the field of stellar astrophysics (\citealt{1966MNRAS.132..347R}; \citealt{1973NPhS..246...93R}). 

We have divided internal magnetic fields into two types: a) purely poloidal, which has non-zero radial and angular components of the magnetic field with zero axial component, and b) purely toroidal, which has a non-zero axial component of the magnetic field but with zero radial and angular components. In this paper, we specialize to the case of a dipolar magnetic field $(l=1)$, whose configuration we estimate analytically, but it should be noted that an equivalent procedure can be formulated for any multipole. 
 
\subsection{Internal Magnetic field configuration}

Following the procedure laid out by \cite{Haskell_2008}, we impose two conditions on the internal magnetic field configuration to determine it completely. The boundary conditions for the magnetic fields are dictated by their behaviour at the star's centre and at infinity. The magnetic field of the star is regular at the center of the star, goes to zero at infinity outside the star and must be continuous with the external magnetic field at the surface of the star. We start off with the purely poloidal magnetic field. The first condition being imposed is the divergence-free condition 
\begin{equation}\label{deldotB}
    \Vec{\nabla}\cdot\Vec{B} = 0
\end{equation} 
which gives us the condition 
\begin{equation}
    \frac{dB_{r}}{dr} + \frac{2B_{r}}{r} - \frac{\beta B_{\perp}}{r} = 0,
\end{equation} 
for the poloidal magnetic fields. The next condition comes from taking a curl of equation \eqref{eq:5}, which gives us the result that the curl of the  Lorentz force is zero. This is the barotropic MHD equilibrium condition, expressed as 
\begin{equation}\label{eq:MHD} \Vec{\nabla}\cross\bigg(\frac{(\Vec{\nabla}\cross\Vec{B})\cross\Vec{B}}{4\pi\rho}\bigg)=0.
\end{equation} which is well defined throughout the entire volume of the neutron star for the polytropic $n=1$ Equation of State. Applying this condition to poloidal magnetic fields, we end up with the equation
\[\Bigg(B_{r} +  r\frac{dB_{r}}{dr} - \frac{rB_{r}}{\rho}\frac{d\rho}{dr} +  \frac{B_{\perp}}{\beta}\Bigg)\Bigg(\frac{dB_{\perp}}{dr} + \frac{B_{\perp}}{r} -   \frac{B_{r}}{r}\Bigg) + \]\begin{align}
rB_{r}\Bigg(\frac{d^{2}B_{\perp}}{dr^{2}} + \frac{1}{r}\frac{dB_{\perp}}{dr} - \frac{B_{\perp}}{r^{2}} + \frac{B_{r}}{r^{2}} - \frac{1}{r}\frac{dB_{r}}{dr}\Bigg)= 0.\label{eq:MHDfull1} 
\end{align}
We obtain the radial and angular components of the dipole $(l=1)$ poloidal magnetic field by using the analytical solutions calculated by \citealt{Haskell_2008} as
\begin{equation}
\label{B_r}
B_{r}(r) = \frac{2F_{0}\rho_{c}}{\pi^{5}}\Bigg[\bigg(\frac{3\pi^{2}R^{3}}{r} - \frac{6R^{5}}{r^{3}}\bigg)\sin \left(\frac{\pi r}{R} \right) +
\frac{6\pi R^{4}}{r^{2}}\cos\left( \frac{\pi r}{R}\right) + \pi^{3}R^{2}\Bigg], 
\end{equation}
\[B_{\perp}(r) = \frac{\beta F_{0}\rho_{c}}{\pi^{5}}\Bigg[\bigg(\frac{3\pi^{2}R^{3}}{r} + \frac{6R^{5}}{r^{3}}\bigg)\sin\left( \frac{\pi r}{R}\right) + \]\begin{equation}\bigg(3\pi^{2}R^{3} -\frac{6R^{5}}{r^{2}}\bigg)\frac{\pi}{R}\cos\left( \frac{\pi r}{R}\right) - \frac{6\pi^{2}R^{3}}{r}\sin\left(\frac{\pi r}{R}\right) + 2\pi^{3}R^{2}\Bigg].    
\end{equation} where $F_{0}$ is a constant determined by the maximum strength of the poloidal magnetic field inside the neutron star. 

Next, we consider a purely toroidal magnetic field. The first divergence-free condition in equation (\ref{deldotB}) is automatically satisfied for the toroidal magnetic fields. \begin{equation}
    \Vec{\nabla}\cdot\bigg(B_{\cross}\frac{r}{\beta}(\hat{r}\cross\Vec{\nabla}Y_{lm})\bigg) = 0.
\end{equation} 
Next, we apply the MHD equilibrium condition in equation (\ref{eq:MHD}) to toroidal magnetic fields and we end up with the equation \begin{equation}\label{eq:MHDfull2} \frac{B^{2}_{\cross}}{\rho}\frac{d\rho}{dr} - B_{\cross}\frac{dB_{\cross}}{dr} = 0.\end{equation} We thus obtain the axial component of the internal toroidal magnetic field \citet{Haskell_2008} as  
\begin{equation}\label{B_cross}
    B_{\cross}(r) = \Big(\frac{4\pi}{3}\Big)^{1/2}\frac{\beta S_{0}\rho_{c}}{\pi^{5}}
    \sin \left(\frac{\pi r}{R}\right). 
\end{equation} where $F_{0}$ is a constant determined by the maximum strength of the toroidal magnetic field inside the neutron star.

For the purely poloidal magnetic fields, the Lorentz force is expressed as 
\begin{equation}
    f_{r} = -\frac{1}{4\pi}\Bigg(\frac{dB_{\perp}}{dr} + \frac{B_{\perp}}{r} - \frac{B_{r}}{r}\Bigg)B_{\perp}\frac{r^{2}}{\beta^{2}}\mathbf{A}_{2}
    \end{equation}
\begin{equation}
    f_{\perp} = \frac{1}{4\pi}\Bigg(\frac{dB_{\perp}}{dr} + \frac{B_{\perp}}{r} - \frac{B_{r}}{r}\Bigg)B_{r}\frac{r}{\beta}\mathbf{B}_{1} 
\end{equation}
and for the purely toroidal magnetic fields they are expressed as \begin{equation}
    f_{r} = -\frac{1}{4\pi}\Bigg(\frac{dB_{\cross}}{dr} + \frac{B_{\cross}}{r}\Bigg)B_{\cross}\frac{r^{2}}{\beta^{2}}\mathbf{A}_{2}
\end{equation}
\begin{equation}
    f_{\perp} =  \frac{1}{4\pi}B^{2}_{\cross}\mathbf{B}_{1}
\end{equation} 
where the coefficients $\mathbf{B}_{1}$ and $\mathbf{A}_{2}$ are obtained after the triple integrals of vector spherical harmonics' components detailed in Appendix \ref{appendix A} and the values of the VSH coupling coefficients are calculated to be $\mathbf{A}_{2} =  \frac{3\pi}{64r^{2}}\sqrt{\frac{5}{\pi}}$ and $\mathbf{B}_{1} = \frac{9\pi}{32r}\sqrt{\frac{5}{\pi}}$ for $(l_{1},m_{1}) = (1,0)$, $(l_{2},m_{2}) = (1,0)$ and $(l_{3},m_{3}) = (2,0)$. 

\subsection{Internal Magnetic field configuration with surface currents}\label{sbs:SC}
Now, following \citet{2014MNRAS.445.2777F}, we introduce the toroidal surface currents at the core-crust boundary $(r=r_c)$ and the crust-ocean boundary $(r=r_o)$ using the expression
\begin{align}
    \Vec{J}_{sur} = K_0(\delta(r - r_c) - \delta(r - r_o))\frac{r}{\beta}[\hat{r}\cross\Vec{\nabla}Y_{lm}].
\end{align}
where $K_{0}$ quantifies the strength of the surface currents. These surface currents generate a purely dipolar magnetic field, which modifies the internal magnetic field of the star while ensuring that they still satisfy the equations \eqref{deldotB}, \eqref{eq:MHDfull1} and \eqref{eq:MHDfull2}. Using equation (D10) in \citet{2014MNRAS.445.2777F}, the modified flux function results in the radial and angular components of the dipolar ($l=1$) poloidal magnetic field being modified as 
\begin{align}
    B_{r}(r) =  
    \left\{
    \begin{aligned}
        & \Big(\frac{4\pi}{3}\Big)^{3/2}\frac{2F_{0}\rho_{c}}{\pi^{5}}\Bigg[\bigg(\frac{3\pi^{2}R^{3}}{r} - \frac{6R^{5}}{r^{3}}\bigg)\sin \left(\frac{\pi r}{R} \right) +\\
        &\frac{6\pi R^{4}}{r^{2}}\cos\left( \frac{\pi r}{R}\right) 
        +\pi^3 R^2 \big], (r \leq r_c) \\
        & \Big(\frac{4\pi}{3}\Big)^{3/2}\frac{2F_{0}\rho_{c}}{\pi^{5}}\Bigg[\bigg(\frac{3\pi^{2}R^{3}}{r} - \frac{6R^{5}}{r^{3}}\bigg)\sin \left(\frac{\pi r}{R} \right) +
        \frac{6\pi R^{4}}{r^{2}}\cos\left( \frac{\pi r}{R}\right) \\  &+\pi^3 R^2 \big] + 2K_{0}\left(\frac{r_c^3}{r^3} - 1 \right), (r_c \leq r \leq r_o) \\
        & \Big(\frac{4\pi}{3}\Big)^{3/2}\frac{2F_{0}\rho_{c}}{\pi^{5}}\Bigg[\bigg(\frac{3\pi^{2}R^{3}}{r} - \frac{6R^{5}}{r^{3}}\bigg)\sin \left(\frac{\pi r}{R} \right) +
        \frac{6\pi R^{4}}{r^{2}}\cos\left( \frac{\pi r}{R}\right)\\     &+\pi^3 R^2 \Bigg] +  2K_{0}\left(\frac{r_c^3}{r^3} - \frac{r_o^3}{r^3} \right) , (r_o \leq r \leq R) 
    \end{aligned}
    \right.
\end{align}
\begin{align}
    B_{\perp}(r) =  
    \left\{
    \begin{aligned}
        & \Big(\frac{4\pi}{3}\Big)^{3/2}\frac{\beta F_{0}\rho_{c}}{\pi^{5}}\Bigg[\bigg(\frac{3\pi^{2}R^{3}}{r} + \frac{6R^{5}}{r^{3}}\bigg)\sin\left( \frac{\pi r}{R}\right) + \\ &\bigg(3\pi^{2}R^{3} -\frac{6R^{5}}{r^{2}}\bigg)\frac{\pi}{R}\cos\left( \frac{\pi r}{R}\right) - \frac{6\pi^{2}R^{3}}{r}\sin\left(\frac{\pi r}{R}\right)  \\
        & + 2\pi^{3}R^{2}\Bigg], (r \leq r_c) \\
        & \Big(\frac{4\pi}{3}\Big)^{3/2}\frac{\beta F_{0}\rho_{c}}{\pi^{5}}\Bigg[\bigg(\frac{3\pi^{2}R^{3}}{r} + \frac{6R^{5}}{r^{3}}\bigg)\sin\left( \frac{\pi r}{R}\right) + \\ &\bigg(3\pi^{2}R^{3} -\frac{6R^{5}}{r^{2}}\bigg)\frac{\pi}{R}\cos\left( \frac{\pi r}{R}\right) - \frac{6\pi^{2}R^{3}}{r}\sin\left(\frac{\pi r}{R}\right)  \\
        & + 2\pi^{3}R^{2}\Bigg] - \beta K_{0}\left(2 + \frac{r_c^3}{r^3}\right), (r_c \leq r \leq r_o) \\
        & \Big(\frac{4\pi}{3}\Big)^{3/2}\frac{\beta F_{0}\rho_{c}}{\pi^{5}}\Bigg[\bigg(\frac{3\pi^{2}R^{3}}{r} + \frac{6R^{5}}{r^{3}}\bigg)\sin\left( \frac{\pi r}{R}\right) + \\ &\bigg(3\pi^{2}R^{3} -\frac{6R^{5}}{r^{2}}\bigg)\frac{\pi}{R}\cos\left( \frac{\pi r}{R}\right) - \frac{6\pi^{2}R^{3}}{r}\sin\left(\frac{\pi r}{R}\right)  \\
        & + 2\pi^{3}R^{2}\Bigg] + \beta K_{0}\left(\frac{r_o^3}{r^3} - \frac{r_c^3}{r^3}\right), (r_o \leq r \leq R)  \\
    \end{aligned}
    \right.
\end{align} Similarly, the axial component of the dipolar ($l=1$) toroidal magnetic field is calculated as 
\begin{align}
    B_{\cross}(r) =  
    \left\{
    \begin{aligned}
        & \Big(\frac{4\pi}{3}\Big)^{1/2}\frac{\beta S_{0}\rho_{c}}{\pi^{5}}\sin \left(\frac{\pi r}{R}\right), (r \leq r_c) \\
        & \Big(\frac{4\pi}{3}\Big)^{1/2}\frac{\beta (S_{0} + K_{1})\rho_{c}}{\pi^{5}}\sin \left(\frac{\pi r}{R}\right) (r_c \leq r \leq r_o)  \\
        & \Big(\frac{4\pi}{3}\Big)^{1/2}\frac{\beta S_{0}\rho_{c}}{\pi^{5}}\sin \left(\frac{\pi r}{R}\right), (r_o \leq r \leq R)  \\
    \end{aligned}
    \right.
\end{align} where $K_{1}$ analogously quantifies the strength of the poloidal surface currents. These internal magnetic field components are then put back into the aforementioned equations for the Lorentz force components to obtain the modified forces that are enhanced inside the crust. 
\section{Perturbation Equations}\label{section:5}
We construct our system of perturbative equations and estimate the size of a mountain formed on the crust of a neutron star due to the effect of its internal magnetic fields. We divide our star into three separate layers: a fluid core, an elastic crust, and a fluid ocean. We choose to include a fluid layer outside the crust since at low densities, the crustal lattice begins to melt (\citealt{Fantina_2020}), and it also simplifies the matching of the gravitational potential to the exterior potential at the surface of the star. The elasticity of the crust and the internal magnetic fields, as mentioned before, are only perturbative effects in our neutron star system. 

\subsection{The Fluid Core and Ocean}\label{sb:5.1}
We first analyze the perturbation equations of the fluid parts of the star. To express the perturbations induced inside the star due to the Lorentz force generated by the internal magnetic fields, we decompose equation \eqref{eq:7} into its radial and angular components and substitute the expression for the Lorentz force into these equations. Before this, we re-introduce and expand the displacement vector in a multipolar decomposition \begin{equation}
    \Vec{\xi} = \xi_{r}Y_{lm}\hat{r} + \frac{r}{\beta}\xi_{\perp}\Vec{\nabla}Y_{lm}
\end{equation} 
where we have ignored the axial displacement vectors since the axial component of the perturbing Lorentz force is zero. 
Since we are only interested in the quadrupolar deformations which can form axisymmetry-breaking mountains, which in turn lead to the emission of CGWs, we focus only on the $(l,m) = (2,2)$ harmonics. 
Following \citet{2021MNRAS.500.5570G} and  \citet{Morales_2022}, we introduce an $(l,m)=(2,2)$ force density of the form
\begin{align}
    f_{i} = f_r(r) Y_{lm}\hat{r} + f_\perp(r) r \nabla_i Y_{lm},
\end{align}
where the $f_r$ and $f_{\perp}$ are the radial functions of the Lorentz force. We now expand the perturbed density using equation \eqref{eq:delrho} as \begin{equation}
    \delta\rho = -\rho\Bigg(\frac{d\xi_{r}}{dr} + \frac{2\xi_{r}}{r} + \frac{\xi_{r}}{\rho}\frac{d\rho}{dr} - \frac{\beta\xi_{\perp}}{r}\Bigg)
\end{equation} and the perturbed pressure is expressed similarly using equation \eqref{eq:delp}. We obtain the following equations: \begin{equation}\label{eq:9}
\rho\frac{\delta\Phi}{r} - \frac{c^{2}_{s}\rho}{r}\Bigg(\frac{d\xi_{r}}{dr} + \frac{2\xi_{r}}{r} + \frac{\xi_{r}}{\rho}\frac{d\rho}{dr} - \frac{\beta^{2}\xi_{\perp}}{r}\Bigg) = f_{\perp},
\end{equation} \[
\rho\frac{d\delta\Phi}{dr} -g\rho\Bigg(\frac{d\xi_{r}}{dr} + \frac{2\xi_{r}}{r} + \frac{\xi_{r}}{\rho}\frac{d\rho}{dr} - \frac{\beta^{2}\xi_{\perp}}{r}\Bigg) - \Bigg(\rho\frac{dc^{2}_{s}}{dr} + c^{2}_{s}\frac{d\rho}{dr}\Bigg)\Bigg(\frac{d\xi_{r}}{dr} + \frac{2\xi_{r}}{r} \]\[+ \frac{\xi_{r}}{\rho}\frac{d\rho}{dr} - \frac{\beta^{2}\xi_{\perp}}{r}\Bigg) - c^{2}_{s}\rho\Bigg(\frac{d^{2}\xi_{r}}{dr^{2}} + \frac{2}{r}\frac{d\xi_{r}}{dr} - \frac{2\xi_{r}}{r^{2}} + \frac{\xi_{r}}{\rho}\frac{d^{2}\rho}{dr^{2}} + \]\begin{equation}\label{eq:10} \frac{1}{\rho}\frac{d\xi_{r}}{dr}\frac{d\rho}{dr} - \frac{\xi_{r}}{\rho^{2}}(\frac{d\rho}{dr})^{2} + \frac{\beta^{2}\xi_{\perp}}{r^{2}} - \frac{\beta^{2}}{r}\frac{d\xi_{\perp}}{dr}\Bigg) = f_{r},
\end{equation} 
where $g$ is the acceleration due to gravity inside the star and is defined as the gradient of the background gravitational potential $g(r) = \frac{d\Phi}{dr}$. These forces are different depending on the spatial configuration of the fields.
The system of equations still has another parameter $\delta\Phi$, which can be expressed by re-arranging the perturbed Poisson's equation as \begin{equation}\label{eq:11}
    \frac{d^{2}\delta\Phi}{dr^{2}} + \frac{2}{r}\frac{d\delta\Phi}{dr} -\frac{\beta^{2}}{r^{2}}\delta\Phi = -4\pi G\rho\Bigg(\frac{d\xi_{r}}{dr} + \frac{2\xi_{r}}{r} + \frac{\xi_{r}}{\rho}\frac{d\rho}{dr} - \frac{\beta^{2}\xi_{\perp}}{r}\Bigg)
\end{equation} These three equations \eqref{eq:9}, \eqref{eq:10} and \eqref{eq:11} are what we must solve to obtain the perturbations in the fluid part of the star, and we need the appropriate boundary conditions to solve them. The first condition is the regularity of the perturbed potential at the centre of the star, which leads to \begin{equation}\label{BC1}
    \delta\Phi(r = 0) = 0,
\end{equation} the regularity of the gradient of the perturbed potential at the center of the star which leads to \begin{equation}\label{BC2}
    \frac{d\delta\Phi}{dr}(r = 0) = 0,
\end{equation} and the second condition is the matching of the perturbed potential at the surface to the external solution, which gives us \begin{equation}\label{BC3}
    \frac{d\delta\Phi}{dr}(r = R) = - \frac{(l+1)}{R}\delta\Phi(r = R).
\end{equation} Our final boundary condition comes from setting the Lagrangian perturbed pressure to zero at the surface of the star $\Delta p(r=R) = 0 \implies \delta p(r = R) = 0$ which gives us the condition \[
    -c^{2}_{s}(r = R)\rho(r = R)\Bigg(\frac{d\xi_{r}}{dr}(r = R) + \frac{2\xi_{r}(r = R)}{R} +\]\begin{equation}\frac{\xi_{r}(r = R)}{\rho(r = R)}\frac{d\rho}{dr}(r = R) - \frac{\beta\xi_{\perp}(r = R)}{R}\Bigg) = 0.
\end{equation} 

\subsection{The Elastic Crust}\label{sb:5.2}
We now analyze the perturbation equations for the elastic crust of the neutron star. An emergent effect that we encounter here is the elasticity of the crust, which enters the system of equations only in the perturbative order. To characterize this elasticity of the crust, we introduce the shear-stress tensor as described in equation \eqref{eq:12}. For the stress tensor $\sigma_{ij}$ (\citealt{Ushomirsky_2000}), we also have \begin{equation}
    t_{ij} = 2\mu\sigma_{ij}.
\end{equation} For ease of calculations, we now introduce the perturbed traction vector, defined here as (\citealt{Haskell_2008}) \[
    T^{i} = (\delta pg^{ij} - t^{ij} - \frac{1}{4\pi}\Big(B^{i}B^{j} - \frac{B^{2}}{2}g^{ij}\Big))\hat{r}_{j} \implies\]\begin{equation}
     \Vec{T} = \Big(\delta p(r) - T_{1}(r) + \frac{B^{2}}{8\pi} - \frac{(B^{r})^{2}}{4\pi}\Big)\hat{r}Y_{lm} - \frac{r}{\beta}\Big(T_{2}(r) + \frac{B^{r}B^{\perp}}{4\pi}\Big)\Vec{\nabla}Y_{lm}
\end{equation} for the purely poloidal magnetic fields and \[
     T^{i} = (\delta pg^{ij} - t^{ij} - \frac{1}{4\pi}\Big(B^{i}B^{j} - \frac{B^{2}}{2}g^{ij}\Big))\hat{r}_{j} \implies\]\begin{equation}
     \Vec{T} = \Big(\delta p(r) - T_{1}(r) + \frac{B^{2}_{\cross}}{8\pi} \Big)\hat{r}Y_{lm} - \frac{r}{\beta}T_{2}(r) \Vec{\nabla}Y_{lm}
\end{equation} for the purely toroidal magnetic fields. Now, the new radial functions $T_{1}$ and $T_{2}$, useful for our calculations, are defined using equation \eqref{eq:12} as \begin{equation}
    T_{1}(r)Y_{lm} \equiv t_{rr} = \frac{2\mu}{3r}\Bigg(2r\frac{d\xi_{r}}{dr} -2\xi_{r} + \beta\xi_{\perp}\Bigg)Y_{lm},
\end{equation}\begin{equation}
    T_{2}(r)\nabla_{a}Y_{lm} \equiv \frac{\beta}{r}t_{ra} = \frac{\mu}{r}\Bigg(r\frac{d\xi_{\perp}}{dr} + \beta\xi_{r} - \xi_{\perp}\Bigg) \nabla_{a}Y_{lm}, 
\end{equation} where $a$ can be either of the angular coordinates $\theta$ or $\phi$. We now formulate our system of equations using these newly defined traction variables. We re-express their definition to obtain \begin{equation}\label{eq:13}
    \frac{d\xi_{r}}{dr} = \frac{\xi_{r}}{r} - \frac{\beta\xi_{\perp}}{2r} + \frac{3T_{1}}{4\mu},
\end{equation} \begin{equation}\label{eq:14}
    \frac{d\xi_{\perp}}{dr} = -\frac{\beta\xi_{r}}{r} + \frac{\xi_{\perp}}{r} + \frac{T_{2}}{\mu}.
\end{equation} 
We repeat the exercise of decomposing the perturbed Euler equation \eqref{eq:5} in the elastic crust into its radial and angular components and insert the Lorentz force into these equations, also using the traction variables this time. We obtain the following equations: \[
    \Bigg(1+\frac{3c^{2}_{s}\rho}{4\mu}\Bigg)\frac{dT_{1}}{dr} = \rho \frac{d\delta\Phi}{dr} + f_{r} -\Bigg(\frac{dc^{2}_{s}}{dr}\Bigg(3\rho + r\frac{d\rho}{dr}\Bigg) + \]\[ c^{2}_{s}\Bigg(\frac{3\beta^{2}\rho}{2r} + \frac{d\rho}{dr} - \frac{r}{\rho}\Bigg(\frac{d\rho}{dr}\Bigg)^{2} + r\frac{d^{2}\rho}{dr^{2}}\Bigg)\Bigg)\frac{1}{r}\xi_{r} + \Bigg(\frac{dc^{2}_{s}}{dr}3\rho + \]\[ c^{2}_{s}\Bigg(\frac{3\rho}{r} + \frac{d\rho}{dr}\Bigg)\Bigg)\frac{\beta}{2r}\xi_{\perp} - \Bigg(\frac{3}{r} + \frac{dc^{2}_{s}}{dr}\frac{3\rho}{4\mu} + c^{2}_{s}\Bigg(\frac{3\rho}{r} -\frac{\rho}{\mu}\frac{d\mu}{dr} +\]\begin{equation}\label{eq:15} \frac{d\rho}{dr}\Bigg)\frac{3}{4\mu}\Bigg)T_{1} + \Bigg(1 + \frac{3c^{2}_{s}\rho}{2\mu}\Bigg)\frac{\beta^{2}}{r}T_{2},
    \end{equation}\[
    \frac{dT_{2}}{dr} = \frac{\rho}{r}\delta\Phi + f_{\perp} - c^{2}_{s}\Bigg(3\rho + r\frac{d\rho}{dr}\Bigg)\frac{1}{r^{2}}\xi_{r} + \Bigg(\frac{3c^{2}_{s}\rho}{2} + \]\begin{equation}\label{eq:16} \Bigg(1 - \frac{2}{\beta^{2}}\Bigg)\mu\Bigg)\frac{\beta}{r^{2}}\xi_{\perp} + \Bigg(\frac{1}{2} - \frac{3c^{2}_{s}\rho}{4\mu}\Bigg)\frac{1}{r}T_{1} - \frac{3}{r}T_{2}.
\end{equation} 
To complete our system of equations, we use the Perturbed Poisson's equation again \eqref{eq:11} along with the appropriate boundary conditions to solve this system of equations, which are borrowed from the previous subsection. The regularity at the centre of the star \eqref{BC1}, \eqref{BC2} and continuity at the surface of the perturbed gravitational potential and its radial gradient \eqref{BC3} are utilised as boundary conditions here. In this case, we also obtain four extra boundary conditions considering the continuity of the perturbed traction vector across the two crustal boundaries. For the purely poloidal magnetic fields, the radial traction vector gives us the continuity of  \begin{equation}\label{T1P}
    \delta p - T_{1} + \frac{(B^{\perp})^{2} - (B^{r})^{2}}{8\pi}
\end{equation} across both the crustal boundaries, with $T_{1}$ vanishing outside the crust. For the angular traction vector, we must have the continuity of \begin{equation}\label{T2P}
    T_{2} - \frac{B^{r}B^{\perp}}{4\pi}
\end{equation} across both the crustal boundaries, with $T_{2}$ vanishing outside the crust. For the purely toroidal magnetic fields, the radial traction vector gives us the continuity of \begin{equation}\label{T1T}
    \delta p - T_{1} + \frac{(B^{\cross})^{2}}{8\pi}
\end{equation} across both the crustal boundaries, with $T_{1}$ vanishing outside the crust. For the angular traction vector, we simply have the condition \begin{equation}\label{T2T}
    T_{2} = 0
\end{equation} across both the crustal boundaries. Note that the magnetic fields in these boundary conditions are continuous throughout the entire volume of the star. We have thus obtained our system of equations \eqref{eq:13}-\eqref{eq:16} for perturbations in the elastic crust with their appropriate boundary conditions that we solve numerically.

In the presence of surface currents, the components of the magnetic field and therefore, the Lorentz force components are modified according to the prescription in subsection \ref{sbs:SC}. The boundary conditions for the traction components at the crustal boundaries are also modified due to discontinuities of the magnetic fields in the presence of crustal surface currents. This discontinuity is enhanced at the core-crust boundary compared to the crust-ocean boundary, which leads to a jump in magnetic field strength inside the crust.
\subsection{Numerical procedure}

In this subsection, we describe the numerical procedure used by us to solve our system of coupled linear perturbative equations for the fluid core and fluid ocean of the star (equations \eqref{eq:9}, \eqref{eq:10} and \eqref{eq:11}), and for the elastic crust of the star (equations \eqref{eq:13}, \eqref{eq:14}, \eqref{eq:15}, \eqref{eq:16} and \eqref{eq:11}) using their respective boundary conditions. First, we describe the emergent properties of our star when it is perturbed; i.e. the elasticity and the magnetic field's spatial configuration of the star. The elastic crust's shear modulus profile is given by equation \eqref{eq:kappa}, where the value of constant $\kappa$ is taken to be $ \kappa = 1.0 \times 10^{16} \text{cm}^{2}\text{s}^{-2}$. Since we estimate the size of the mountain when the crust of the star is maximally strained, we calculate the von Mises strain $\overline{\sigma}$. The von Mises strain is defined using the strain tensor (\citealt{Ushomirsky_2000}): 
\begin{equation}
    \overline{\sigma}^{2} = \frac{1}{2}\sigma_{ij}\sigma^{ij}.
\end{equation} The von Mises criterion states that an elastic material will reach its yield limit when $\overline{\sigma} \geq \overline{\sigma}_{max}$. For quadrupolar ($(l,m) = (2,2)$) perturbations, we obtain (\citealt{2021MNRAS.500.5570G})\[
    \overline{\sigma}^{2} = \frac{5}{256\pi}\Bigg[6\sin^{2}{\theta} \Bigg[3\sin^{2}{\theta}\cos^{2}{2\phi}\bigg(\frac{T_1}{\mu}\bigg)^{2} + 4(3 + \cos{2\theta} - 2\sin^{2}{\theta}\cos{4\phi}) \]\begin{equation}\label{vonMises} \times\bigg(\frac{T_2}{\mu}\bigg)^{2}\Bigg] + (35 + 28\cos{2\theta} + \cos{4\theta} + 8\sin^{4}{\theta}\cos{4\phi}) \bigg(\frac{\xi_{\perp}}{r}\bigg)^{2}\Bigg]
\end{equation}
Since the von Mises strain is a function of position, we can identify the location where it is the highest and this is the point where the crust will be ruptured first. We assume this breaking strain to be $\overline{\sigma}_{max} = 10^{-1}$ (\citealt{Horowitz_2009}). 


The numerical procedure to solve this system of equations is the initial value method implemented by using the \texttt{Vern7} solver, which is part of the \texttt{Julia} programming language's \texttt{DifferentialEquations} package. It is a seventh-order explicit Runge-Kutta method which is often used for solving linear coupled ordinary differential equations. 
The perturbations are then normalized by ensuring that the exact point in the crust where the strain is maximum reaches breaking strain first according to equation \eqref{vonMises}. This normalization of the Lorentz force is done by rescaling the strength of the magnetic field (and the crustal surface currents when they are present) inside the star. This condition acts as the convergence criterion for our iterative numerical procedure. 

\section{Numerical results}\label{section:6}
\subsection{Internal magnetic fields}
\begin{figure}
    \centering
    \includegraphics[width=10cm]{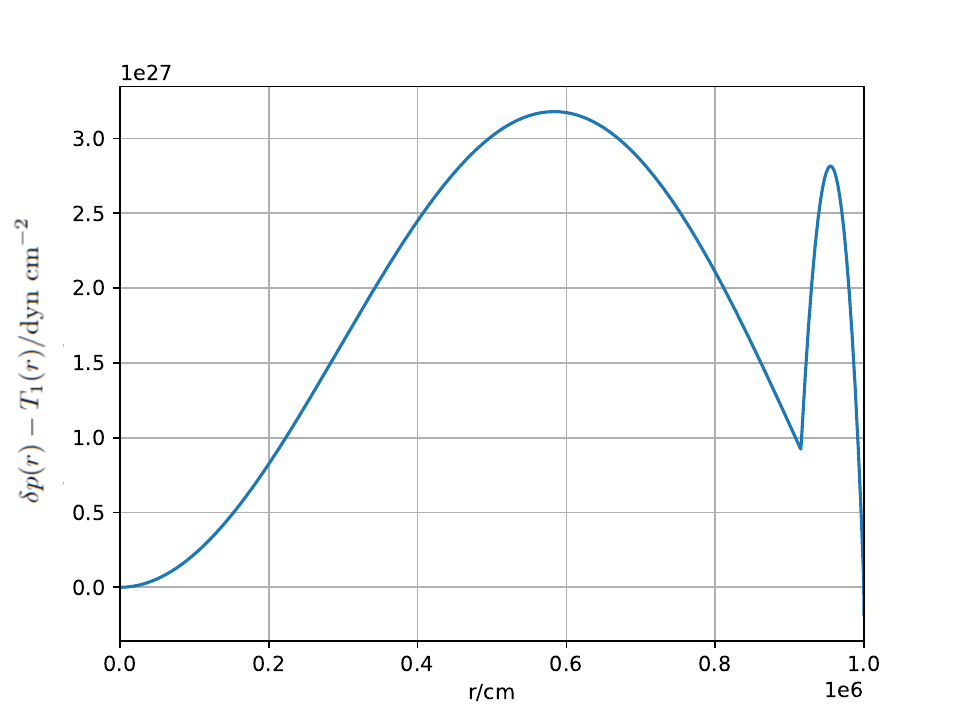}
    \caption{Poloidal case: The perturbed radial Traction component ($\delta p - T_{1}$)}
    \label{fig:3}
\end{figure}

\begin{figure}
    \centering
    \includegraphics[width=10cm]{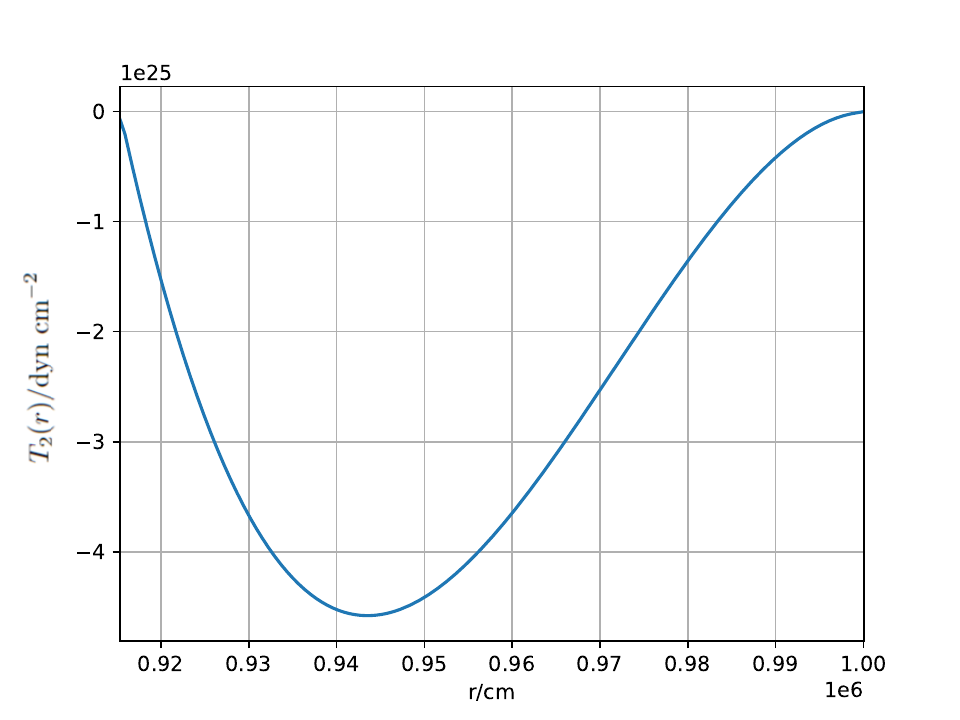}
    \caption{Poloidal case: The perturbed angular Traction component ($T_{2}$)}
    \label{fig:4}
\end{figure}

\begin{figure}
    \centering
    \includegraphics[width=10cm]{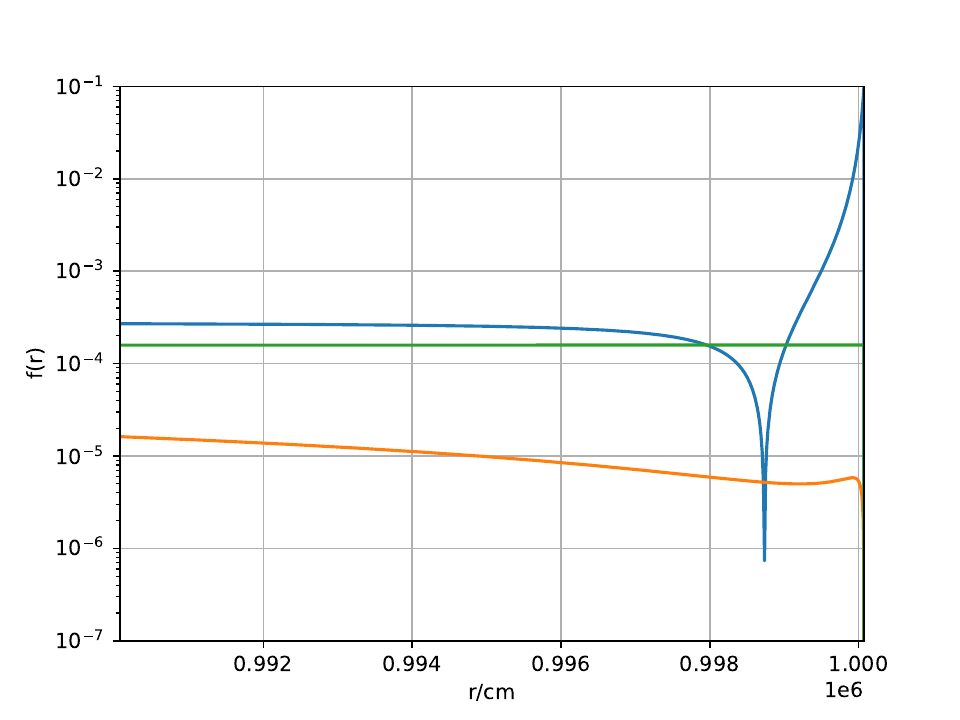}
    \caption{Poloidal case: The strain components maximized over ($\theta,\phi$) against radius at the point where the crust is broken. The individual plots and their scaling are explained in the text.}
    \label{fig:5}
\end{figure}

\begin{figure}
    \centering
    \includegraphics[width=10cm]{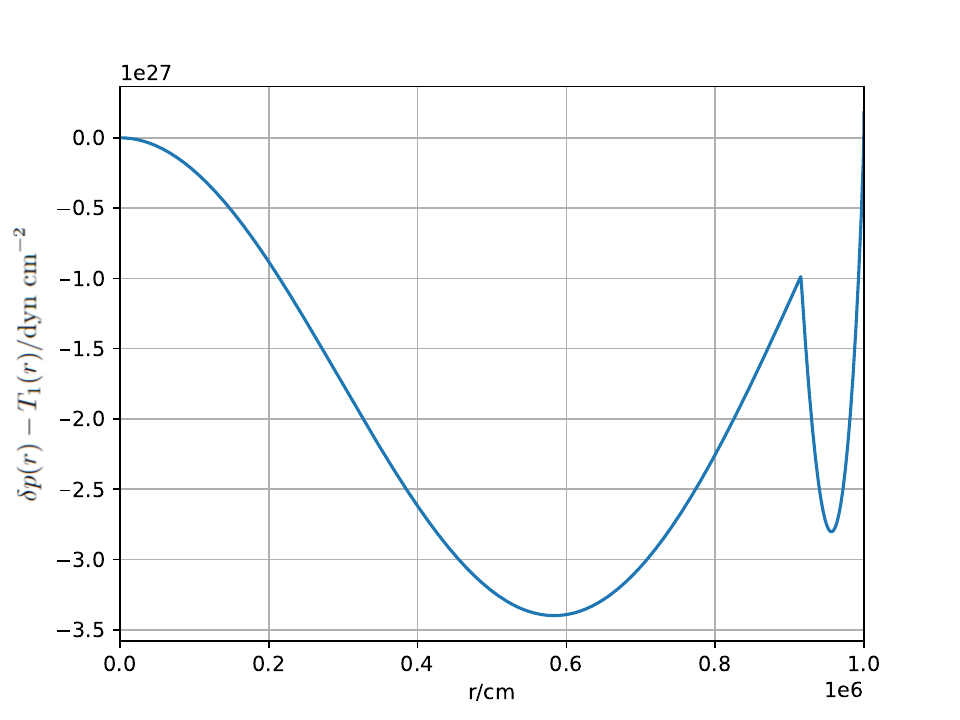}
    \caption{Toroidal case: The perturbed radial Traction component ($\delta p - T_{1}$)}
    \label{fig:6}
\end{figure}

\begin{figure}
    \centering
    \includegraphics[width=10cm]{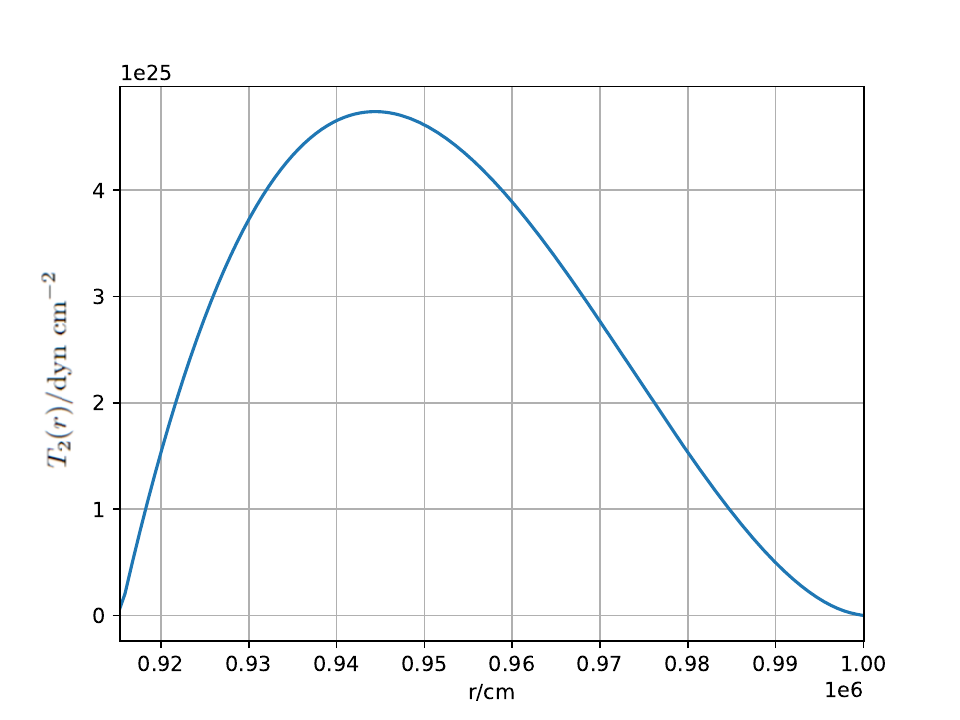}
    \caption{Toroidal case: The perturbed angular Traction component ($T_{2}$)}
    \label{fig:7}
\end{figure}

\begin{figure}
    \centering
    \includegraphics[width=10cm]{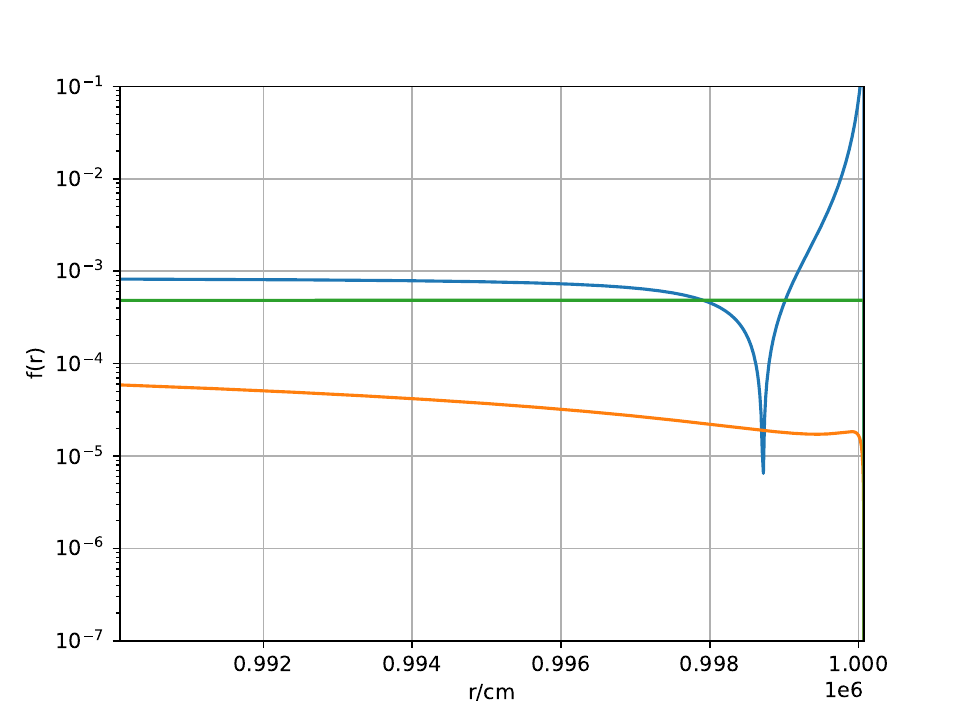}
    \caption{Toroidal case: The strain components maximized over ($\theta,\phi$) against radius at the point where the crust is broken. The individual plots and their scaling are explained in the text.}
    \label{fig:8}
\end{figure}

For the purely poloidal case, Figs. \ref{fig:3} and \ref{fig:4} demonstrate that the perturbed radial traction denoted by equation \eqref{T1P} and the perturbed angular traction denoted by equation \eqref{T2P} are continuous at the fluid-elastic interfaces. We observe that Fig. \ref{fig:5} shows how the dominant contribution to the von Mises strain expression comes from the radial traction component. In Figs. \ref{fig:5} and \ref{fig:8}, the parameters $f(r)$ are expressed as \begin{equation}
    f(r) = \sqrt{\frac{45}{128\pi}}\abs{\frac{T_{1}(r)}{\mu(r)}},
\end{equation} denoted by the blue-coloured line  \begin{equation}
    f(r) = \sqrt{\frac{1215}{512\pi}}\abs{\frac{T_{2}(r)}{\mu(r)}},
\end{equation} denoted by the green-coloured line \begin{equation}
    f(r) = \sqrt{\frac{5}{4\pi}}\abs{\frac{\xi_{\perp}(r)}{r}},
\end{equation} denoted by the red-coloured line. For the purely toroidal case, Figs. \ref{fig:6} and \ref{fig:7} demonstrate that the perturbed radial traction denoted by equation \eqref{T1T} and the perturbed angular traction denoted by equation \eqref{T2T} are continuous at the fluid-elastic interfaces. Similarly, we observe that Fig. \ref{fig:8} shows how the dominant contribution to the von Mises strain expression also comes from the radial traction component. The crust breaks at the very top, where the shear modulus $\mu$ is of the least magnitude, and the star is the weakest at the top for the $(l,m) = (2,2)$ mode. The quadrupole moments are then calculated using equations \eqref{quadrupoles} and \eqref{eq:delrho}, with the corresponding fiducial ellipticities being calculated using equation \eqref{ellip}. This exercise is done for the relaxed star to obtain $Q_{relax}$ and $\epsilon_{relax}$, for the elastically strained star to obtain $Q_{strain}$ and $\epsilon_{strain}$, and then the difference $\abs{Q_{strain} - Q_{relax}}$ and $\abs{\epsilon_{strain} - \epsilon_{relax}}$ gives us the estimate for the maximum size of the \emph{magnetic mountains}. This numerical procedure is done for both purely poloidal and purely toroidal magnetic fields, and the calculated values are displayed in Table \ref{tab:tab1}. 

\begin{table*}
    \centering
    \label{tab:tab1}
    \caption{The maximum quadrupole moments and their corresponding maximum ellipticities calculated for Poloidal and Toroidal magnetic fields}
    \begin{tabular}{ccccc}
    \hline
        Source & $Q_{relax}$ (in $\text{g cm}^{2}$) & $|Q_{strain} - Q_{relax}|$ (in $\text{g cm}^{2}$) & $\epsilon_{relax}$ & $|\epsilon_{strain} - \epsilon_{relax}|$  \\
        \hline
        Poloidal fields & $1.16811 \cross 10^{34}$ & $2.19556 \cross 10^{37}$ & $2.67286 \cross 10^{-11}$ & $1.60856 \cross 10^{-8}$
        \\
        Toroidal fields & $9.28216 \cross 10^{33}$ & $1.18493 \cross 10^{37}$ & $1.21144 \cross 10^{-11}$ & $2.84366 \cross 10^{-8}$
        \\
        \hline
    \end{tabular}
\end{table*}
We see that when the force deforming an elastic star is the Lorentz force generated by the internal magnetic fields of the star, the sizes of the magnetic mountains (quantified by $\abs{\epsilon_{relax} - \epsilon_{strain}}$ and/or $\abs{Q_{relax} - Q_{strain}}$) formed atop the crustal surface is of the same range as predicted by \cite{2021MNRAS.500.5570G} for the case of both the purely poloidal and purely toroidal magnetic fields. As explained by \cite{2021MNRAS.500.5570G}, $\epsilon_{relax}$ and $\abs{\epsilon_{strain} - \epsilon_{relax}}$ are of very different magnitudes because $\epsilon_{relax}$ corresponds to star state \textbf{SM} whose deformation is supported by the Lorentz force with a size limited only by the crustal breaking strain (\citealt{Horowitz_2009}) and the (non-zero) shear modulus of the crust plays little role in this state. In the case of $\abs{\epsilon_{strain} - \epsilon_{relax}}$,  this fiducial ellipticities' difference estimate, which is the indicator of the size of the neutron star mountain, is supported by the shear strains of the crust when the Lorentz force has decayed away, and thus, is sensitive to the shear modulus profile of the crust. However, unlike the results of \cite{2021MNRAS.500.5570G}, the values of $\epsilon_{relax}$ and $Q_{relax}$ are exceptionally smaller than the previous results obtained by \cite{2021MNRAS.500.5570G} and \cite{Morales_2022}. The explanation behind this is a bit subtle. When we estimate the quadrupolar moment $Q$ or the fiducial ellipticity $\epsilon$ for an elastic star state \textbf{SCM} and solve the system of perturbative equations, the radial force is the major contributor to the deformation of the star whereas for the fluid star state \textbf{SM}, the angular force is the major factor which deforms the star. From the angular component of equation \eqref{eq:7}, we see that \begin{equation}
    \delta\rho = \frac{1}{c^{2}_{s}}(f_{\perp} - \rho\delta\Phi),
\end{equation} which can be solved solely with the perturbed Poisson equation \eqref{eq:8} to calculate the quadrupolar moment and thereafter, the fiducial ellipticity. For the Lorentz force, which is a force generated by a solenoidal field, the radial component of the force generated is of similar magnitude compared to the angular component of the force inside the crust, but the modified radial dependence of these forces causes $\epsilon_{strain}$ to be much higher compared to $\epsilon_{relax}$. The radial force is now the major contributor to breaking the crust (as seen in Figs. \ref{fig:5} and \ref{fig:8}), which in tandem with the modified boundary conditions of the traction vector, causes an enhanced deformation of the crust for a magnetic field strength normalized to a lower magnitude. Note that these are the forces generated by magnetic fields of the peak strength of $\sim 10^{15} \text{G}$ (\citealt{Goldreich_Reisenegger_1992}) before they are scaled down to satisfy the maximized straining of one point in the elastic crust. After this magnetic field strength is scaled down, the crust is still strained close to its limit of $\overline{\sigma}_{max} = 10^{-1}$ (\citealt{Horowitz_2009}) by the internal Lorentz forces, but this same magnetic field strength doesn't induce a significant ellipticity for the fluid stars. The elastic stars' ellipticity results or the size of the magnetic mountains are close to what has been observed through gravitational wave observations (\citealt{abbott2020gravitational}, \citealt{abbott2022search}, \citealt{abbott2022searches}, \citealt{donofrio2023search}) and pulsar timing observations (\citealt{falxa2023searching}). We also observe no significant changes in the ellipticity values when we switch from a purely poloidal field to a purely toroidal one, or vice-versa. 

We also observe that the radial traction has a sharp/rapid transition at the core-crust boundary, which is expected given that the traction vector components are non-zero only inside the crust. For the plots in Figs. \ref{fig:5} and \ref{fig:8}, we assumed that the breaking strain of the crust is $\overline{\sigma}_{max} = 10^{-1}$ (\citealt{Horowitz_2009}) and we scale the magnetic fields and corresponding forces accordingly. We observe that similar to the results obtained by \cite{2021MNRAS.500.5570G}, the rupture of the crust is mostly contributed by the radial traction vector $T_{1}$, followed by the angular component of the displacement vector $\xi_{\perp}$ with the angular traction vector $T_{2}$ contributing the least. Thus, dipolar magnetic fields of internal origin in a star do not induce a significantly greater ellipticity in the star before inevitably rupturing the crust. Thus, we have obtained the estimates of the size of the \emph{magnetic mountains} generated by dipolar internal magnetic fields, which are either purely poloidal or purely toroidal. 

\subsection{Internal magnetic fields and Surface currents}
\begin{figure}
    \centering
    \includegraphics[width=10cm]{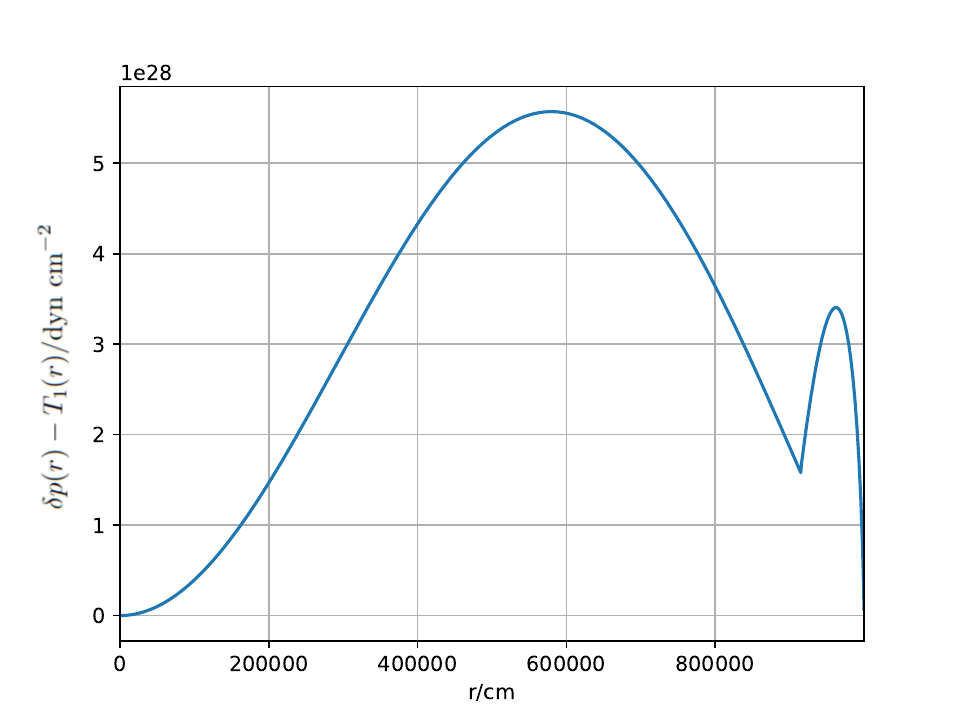}
    \caption{Poloidal case: The perturbed radial Traction component ($\delta p - T_{1}$)}
    \label{fig:9}
\end{figure}

\begin{figure}
    \centering
    \includegraphics[width=10cm]{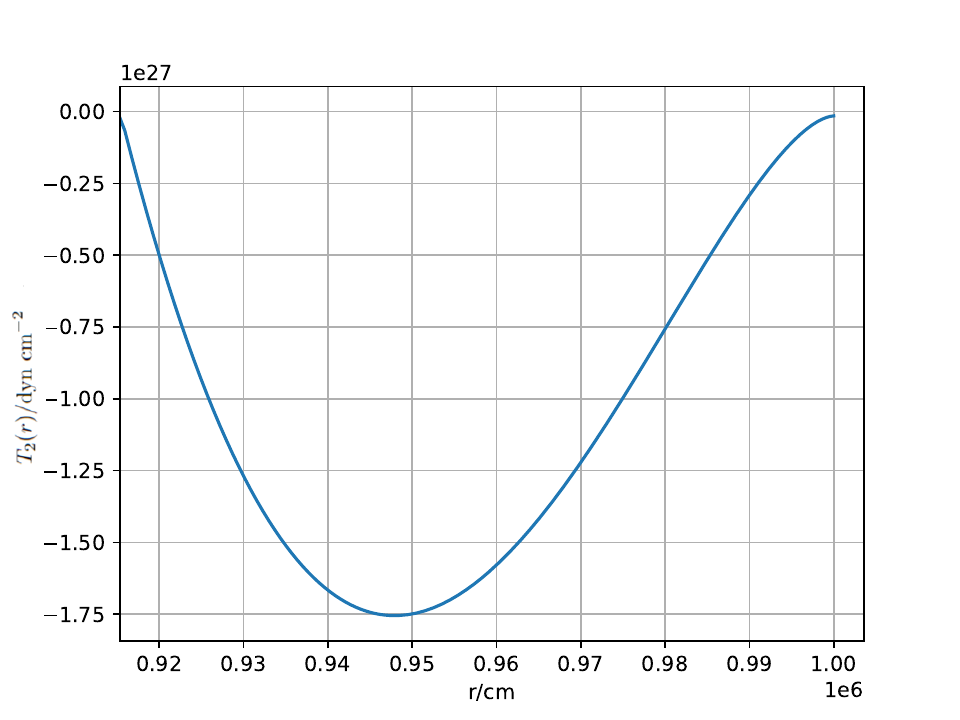}
    \caption{Poloidal case: The perturbed angular Traction component ($T_{2}$)}
    \label{fig:10}
\end{figure}

\begin{figure}
    \centering
    \includegraphics[width=10cm]{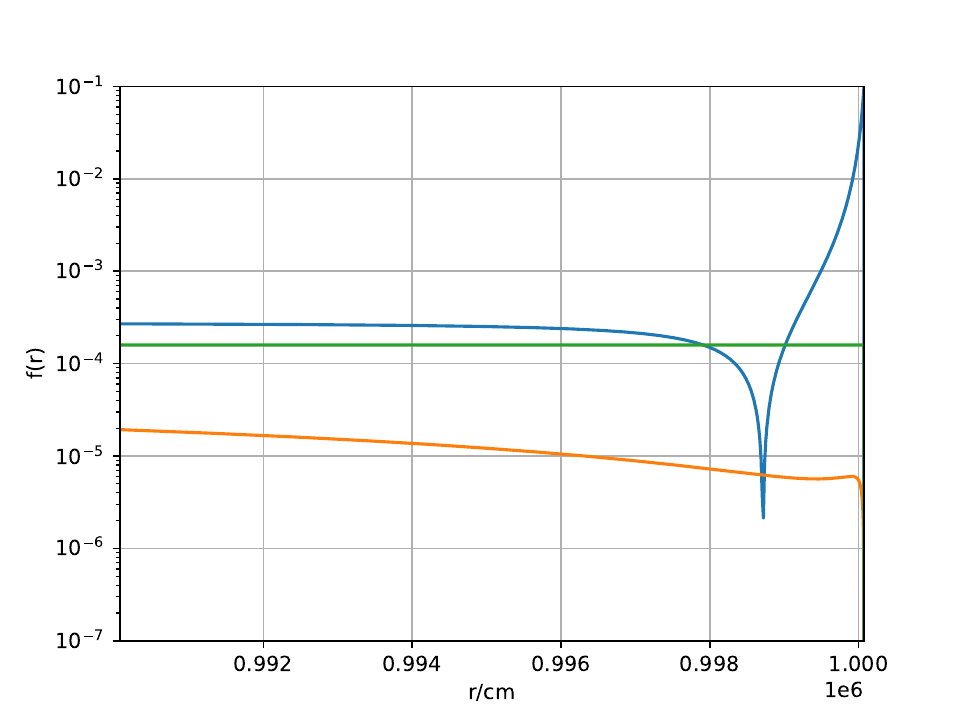}
    \caption{Poloidal case: The strain components maximized over ($\theta,\phi$) against radius at the point where the crust is broken. The individual plots and their scaling are explained in the text.}
    \label{fig:11}
\end{figure}

\begin{figure}
    \centering
    \includegraphics[width=10cm]{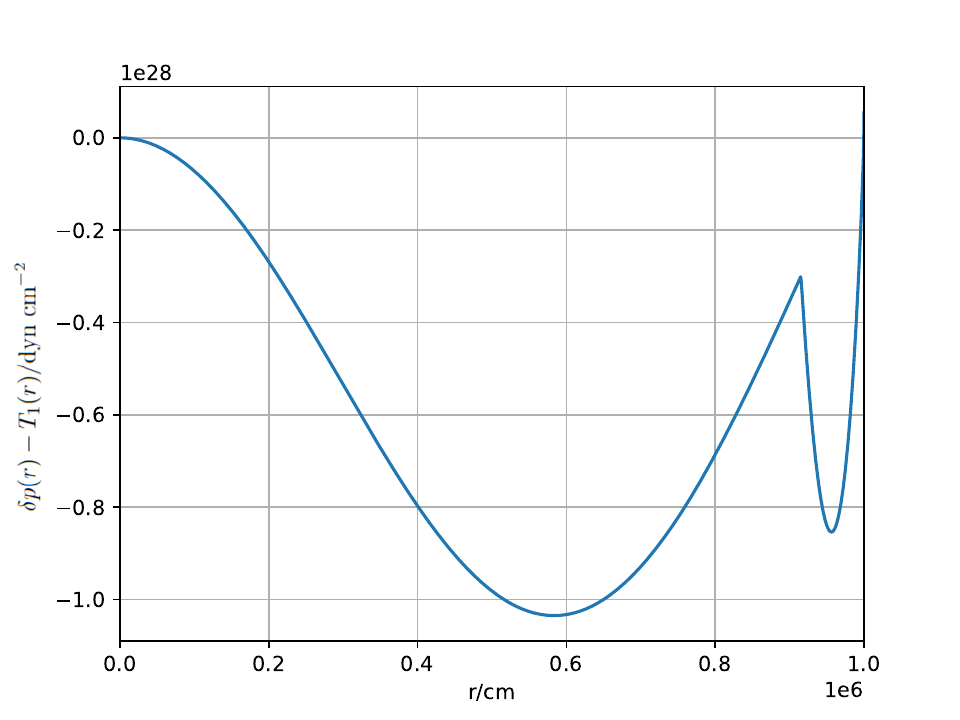}
    \caption{Toroidal case: The perturbed radial Traction component ($\delta p - T_{1}$)}
    \label{fig:12}
\end{figure}

\begin{figure}
    \centering
    \includegraphics[width=10cm]{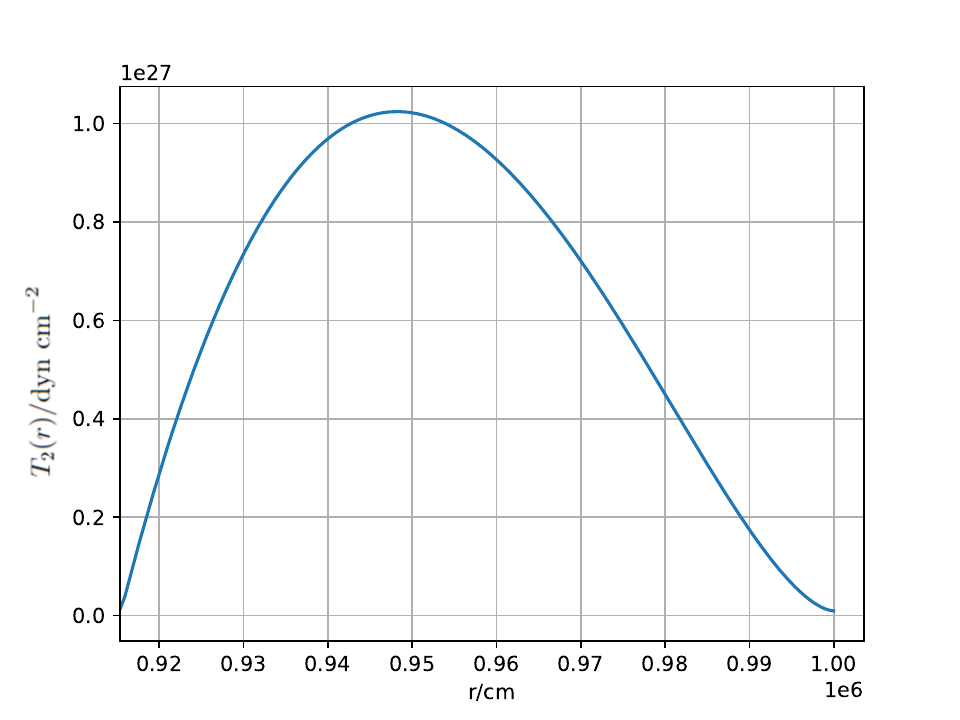}
    \caption{Toroidal case: The perturbed angular Traction component ($T_{2}$)}
    \label{fig:13}
\end{figure}

\begin{figure}
    \centering
    \includegraphics[width=10cm]{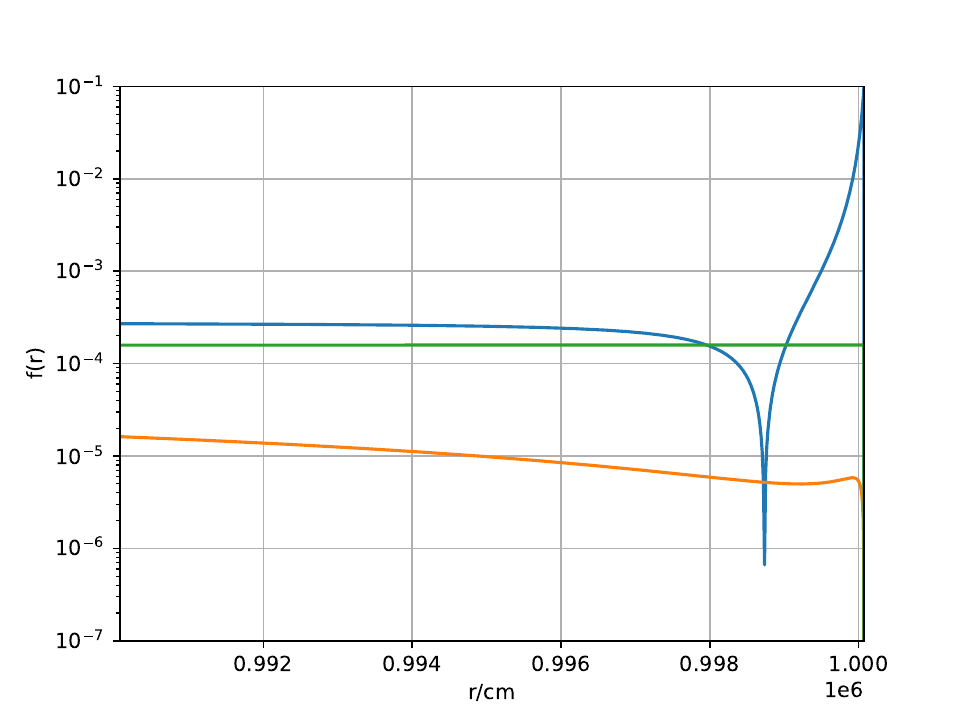}
    \caption{Toroidal case: The strain components maximized over ($\theta,\phi$) against radius at the point where the crust is broken. The individual plots and their scaling are explained in the text.}
    \label{fig:14}
\end{figure}

We now consider the results of the case where the internal magnetic fields are modified by the crustal surface currents. For the purely poloidal case, Figs. \ref{fig:9} and \ref{fig:10} demonstrate that the perturbed radial traction denoted by equation \eqref{T1P} and the perturbed angular traction denoted by equation \eqref{T2P} are continuous at the fluid-elastic interfaces. We observe that Fig. \ref{fig:11} shows how the dominant contribution to the von Mises strain expression comes from the radial traction component. For the purely toroidal case, Figs. \ref{fig:12} and \ref{fig:13} demonstrate that the perturbed radial traction denoted by equation \eqref{T1T} and the perturbed angular traction denoted by equation \eqref{T2T} are continuous at the fluid-elastic interfaces. Similarly, we observe that Fig. \ref{fig:14} shows how the dominant contribution to the von Mises strain expression also comes from the radial traction component. The crust breaks at the very top, where the shear modulus $\mu$ is of the least magnitude, and the star is the weakest at the top for the $(l,m) = (2,2)$ mode. The quadrupole moments are then calculated using equations \eqref{quadrupoles} and \eqref{eq:delrho}, with the corresponding fiducial ellipticities being calculated using equation \eqref{ellip}. This exercise is done for the relaxed star to obtain $Q_{relax}$ and $\epsilon_{relax}$, for the elastically strained star to obtain $Q_{strain}$ and $\epsilon_{strain}$, and then the difference $\abs{Q_{strain} - Q_{relax}}$ and $\abs{\epsilon_{strain} - \epsilon_{relax}}$ gives us the estimate for the maximum size of the \emph{magnetic mountains}, now with the crustal surface currents also present. This numerical procedure is done for both the poloidal and toroidal cases, and the calculated values are displayed in Table \ref{tab:tab2}. 

\begin{table*}
    \centering
    \label{tab:tab2}
    \caption{The maximum quadrupole moments and their corresponding maximum ellipticities calculated for Poloidal and Toroidal magnetic fields in the presence of crustal surface currents}
    \begin{tabular}{ccccc}
    \hline
        Source & $Q_{relax}$ (in $\text{g cm}^{2}$) & $|Q_{strain} - Q_{relax}|$ (in $\text{g cm}^{2}$) & $\epsilon_{relax}$ & $|\epsilon_{strain} - \epsilon_{relax}|$  \\
        \hline
        Poloidal fields & $2.27840 \cross 10^{35}$ & $8.46372 \cross 10^{38}$ & $3.45652 \cross 10^{-10}$ & $1.88503 \cross 10^{-6}$
        \\
        Toroidal fields & $7.95513 \cross 10^{35}$ & $3.21778 \cross 10^{38}$ & $4.87986 \cross 10^{-10}$ & $8.67734 \cross 10^{-7}$
        \\
        \hline
    \end{tabular}
\end{table*}
We see that when the force deforming an elastic star is the Lorentz force generated by the internal magnetic fields of the star, the sizes of the magnetic mountains (quantified by $\abs{\epsilon_{relax} - \epsilon_{strain}}$ and/or $\abs{Q_{relax} - Q_{strain}}$) formed atop the crustal surface is of significantly greater than the values predicted by \cite{2021MNRAS.500.5570G} as well as section \ref{section:5} for the case of both the purely poloidal and purely toroidal magnetic fields after including the effect of crustal surface currents. The explanation of the disparity in the values of $\epsilon_{relax}$ and $\abs{\epsilon_{relax} - \epsilon_{strain}}$ remains unchanged from the previous subsection. Note that these are the forces generated by magnetic fields and surface currents with the peak strength of $\sim 10^{15} \text{G}$ and $\sim 10^{14} \text{G}$ respectively before they are scaled down to satisfy the maximized straining of one point in the elastic crust. This scaling also holds for the fluid stars, despite the fact that they do not have well-defined crusts, and the name 'crustal' surface currents won't make sense for them. The elastic stars' ellipticity results or the size of the magnetic mountains are now enhanced compared to observational estimates but are still within the range of values predicted by \cite{2021MNRAS.500.5570G} and \cite{Morales_2022} for different formulations of the deforming force. Similar to the previous case, we observe no significant changes in the ellipticity values when we switch from a purely poloidal field to a purely toroidal one, or vice-versa. 

We again observe that the radial traction has a sharp/rapid transition at the core-crust boundary, which is expected given that the traction vector components are non-zero only inside the crust. The ellipticity estimate for the poloidal magnetic field with a toroidal surface current in the crust is the largest result we have obtained and is thus the most promising case for us. The continuity (or the lack thereof) of the terms contributed by the magnetic fields in the traction vector is governed by the equation \begin{equation}
    \hat{r}\cross[\Vec{B}_{core} - \Vec{B}_{crust}] = \Vec{J}_{sur}
\end{equation} across the crustal boundaries. For the plots in Figs. \ref{fig:11} and \ref{fig:14}, we assumed that the breaking strain of the crust is $\overline{\sigma}_{max} = 10^{-1}$ (\citealt{Horowitz_2009}) and we scale the magnetic fields, surface currents and corresponding forces accordingly. For Figs. \ref{fig:11} and \ref{fig:14}, we observe that similar to the results obtained by \cite{2021MNRAS.500.5570G}, the rupture of the crust is mostly contributed by the radial traction vector $T_{1}$, followed by the angular component of the displacement vector $\xi_{\perp}$ with the angular traction vector $T_{2}$ contributing the least, which is anomalously non-zero due to the modified boundary conditions. Thus, dipolar magnetic fields of internal origin in a star, when combined with crustal surface currents, generate a significantly greater ellipticity in the star before inevitably rupturing the crust. Thus, we have obtained the estimates of the size of the \emph{magnetic mountains} generated by dipolar internal magnetic fields and dipolar crustal surface currents. 

\section{Discussion and conclusion}\label{section:7}

The deformation of neutron stars due to their internal magnetic fields leading to the formation of a \emph{magnetic mountain} is an interesting phenomenon. Estimating the maximum size of such a \emph{mountain} is important for imposing constraints on the strength of CGW emission from rotating neutron stars as well as determining the maximum spin limit that these systems can attain. We chose to venture into this problem with the aim of providing an enhanced size of mountains, parametrized by the star's fiducial ellipticity and quadrupolar moment, generated by its strong internal fields or with assumed crustal surface currents. The results have been mixed. 

We motivate the evolutionary deformation of mountain formation and build our system of linear perturbative coupled ODEs using the Gittins scheme of mountain formation. Our system of equations is driven by the Lorentz force generated by the internal magnetic fields of the star. We divide the internal magnetic fields into purely poloidal and purely toroidal fields and determine their respective analytic forms using the MHD equilibrium condition of the star. We have also assumed a multipolar decomposition of the magnetic fields to simplify our calculations when we input them into our system of ODEs. We choose a dipolar ($(l,m) = (1,0)$) magnetic field generating a quadrupolar ($(l,m) = (1,0)$) Lorentz force using a prescription of VSH integrals and the Wigner-$\mathcal{D}$ function. We set the peak magnitude of these fields by assuming that the elastic crust is maximally strained at one single point. We repeat this entire calculation with assumed crustal surface currents which drastically increase the relative magnitude of the fields inside the crust compared to the rest of the star. This formulation changes the functional form of the magnetic fields (and thereby the Lorentz force generated by these fields) significantly along with the boundary conditions of our system at the crustal boundaries. After finalizing our system of perturbative equations, we solve them using a 6th-order Runge Kutta method and then, plot the perturbed traction components to demonstrate their radial variation, continuity and behaviour at the crustal boundaries. The modifications to the traction vector components and their corresponding boundary conditions due to the surface currents are also demonstrated by the plots. We also demonstrate that the traction components reach the maximum strain of $\overline{\sigma}_{max} = 10^{-1}$ at the top of the crust or the crust-ocean boundary. We completed our calculations by calculating the fiducial ellipticity estimates and quadrupolar moment estimates for relaxed (\textbf{SM}) and strained (\textbf{SCM}) states of the star, finally estimating the size of the \emph{magnetic mountains} of the star. 

We observe that the values of ellipticity calculated by us generated by the internal magnetic fields of the star are consistent with previous theoretical studies as well as within the upper limits set by astrophysical observations. They are significantly enhanced only when we include surface currents in our calculations, which enhance the strength of the magnetic fields while maximally straining the elastic crust. These results have a number of implications. We do not observe a significant change in the values of ellipticity of the star when we switch from a purely poloidal field to a purely toroidal field, or vice-versa. This indicates that for a dipolar magnetic field inside the star, either of these angular configurations will suffice to generate a \emph{magnetic mountain} atop the crust of the star. However, the radial configuration of these magnetic fields is quite important while calculating the ellipticity of the elastically strained crust. This is yet another modification we introduce by including the surface currents at the crustal boundaries, which also enhance the strength of the generated Lorentz force significantly; with the constraint of maximally straining one point in the elastic crust of the neutron star. The surface currents thus enable us to bypass this aforementioned restriction of maximally straining a single point of the crust which significantly scales down the strength of the internal magnetic fields. The Lorentz force generated in such a system is thus equivalent to the step-function force formulated by \cite{Morales_2022}. Thus, the radial dependence of the force, or its radial configuration, along with the modified boundary conditions, play a much more significant role in inducing higher values of ellipticity of the star and enhancing the size of \emph{magnetic mountains}. This has yet another implication; if and when CGWs are detected and the ellipticity constraints' of the neutron star exceed the current upper limit, a very plausible explanation might also be the presence of strong internal magnetic fields with a radial configuration resembling the model used by us. The possibility of the existence of surface currents might also emerge as another phenomenological explanation for the enhanced ellipticity of the neutron star. However, the maximum size of the \emph{magnetic mountains} calculated by us for either of the cases are to be interpreted as a plausible, physical upper limit only.  

There is tremendous scope for improvement of the formalism used and the results obtained in this paper to enhance the size of the \emph{magnetic mountain} of neutron stars. We detail a few such strategies for calculating the enhanced ellipticities of \emph{magnetic mountains}, some of which are works in progress for a future publication. Instead of considering purely poloidal or purely toroidal magnetic fields, we can consider a mixed magnetic field, denoted by equation \eqref{eq:4}. There are two reasons why this formulation is expected to result in increased ellipticity of a neutron star; we can sustain increased magnitudes of the magnetic fields without rupturing the crust and there will be an additional component of the force introduced in this system. This component is in the axial direction and results only due to the mixing of poloidal and toroidal magnetic fields and is expected to generate axial traction in the crust, which will result in a system of 7 coupled perturbative ODEs and the von Mises strain gaining two additional degrees of freedom. Yet another possibility is the deformation of a neutron star due to external magnetic fields of extraordinary strength, which are essentially poloidal vacuum fields. Yet another possibility is to modify the functional forms of the surface currents present at the crustal boundaries from Dirac-delta functions to approximate sharp Gaussians. The case of considering MED equilibrium is a bit subtle (as detailed in Appendix \ref{MED appendix}) and our MHD assumption can possibly be relaxed to the MED equilibrium, which leads to the coupling of the spatial configuration of the magnetic fields and other perturbations inside the crust. It will be interesting to see if such an assumption can lead to increased ellipticity of the neutron star (\citealt{Kojima_2022}; \citealt{Kojima_2023}). The internal magnetic fields can themselves be modified; we can consider special configurations of the magnetosphere of the neutron star by considering higher multipolar fields (\citealt{Kalapotharakos_2021}), assuming a hot neutron star and analyzing the effect of temperature on the magnetic fields (\citealt{yadav2022thermal}), considering a superconducting core (\citealt{bagchi2023qcd}), considering the effect of superfluidity in the star(\citealt{Haskell_2022}), or considering the effect of plasticity in the crust (\citealt{PhysRevLett.24.1191}; \citealt{10.1111/j.1745-3933.2010.00903.x}). 

The most natural possible extension of this work is to consider the effect of internal magnetic fields to generate a \emph{magnetic mountain} in a relativistic framework, following in the footsteps of \cite{Gittins_2021}. This formulation would require us to solve the perturbed Einstein's equation with a modified energy-momentum tensor that also includes the internal magnetic fields of the star. This formulation can be made even more robust by considering a realistic EoS of the neutron star.   

\section*{Acknowledgements}
AN would like to acknowledge Prof. N. A. Andersson, Fabian Gittins, Prof. Umin Lee, Prof. Shota Kisaka, and Prof. Yasufumi Kojima for insightful discussions on this topic. AN is grateful to Prof. Toshikazu Shigeyama and RESCEU for travel support to Nikhef, the University of Amsterdam, and expresses appreciation to the MMCW workshop 2023 organizers for their hospitality and to all conference participants for insightful discussions.

\section*{Data Availability}
Additional data underlying this article will be shared on reasonable request to the corresponding author.



\bibliographystyle{mnras}
\bibliography{unmain}

\begin{thebibliography}{}
\makeatletter
\relax
\def\mn@urlcharsother{\let\do\@makeother \do\$\do\&\do\#\do\^\do\_\do\%\do\~}
\def\mn@doi{\begingroup\mn@urlcharsother \@ifnextchar [ {\mn@doi@} {\mn@doi@[]}}
\def\mn@doi@[#1]#2{\def\@tempa{#1}\ifx\@tempa\@empty \href {http://dx.doi.org/#2} {doi:#2}\else \href {http://dx.doi.org/#2} {#1}\fi \endgroup}
\def\mn@eprint#1#2{\mn@eprint@#1:#2::\@nil}
\def\mn@eprint@arXiv#1{\href {http://arxiv.org/abs/#1} {{\tt arXiv:#1}}}
\def\mn@eprint@dblp#1{\href {http://dblp.uni-trier.de/rec/bibtex/#1.xml} {dblp:#1}}
\def\mn@eprint@#1:#2:#3:#4\@nil{\def\@tempa {#1}\def\@tempb {#2}\def\@tempc {#3}\ifx \@tempc \@empty \let \@tempc \@tempb \let \@tempb \@tempa \fi \ifx \@tempb \@empty \def\@tempb {arXiv}\fi \@ifundefined {mn@eprint@\@tempb}{\@tempb:\@tempc}{\expandafter \expandafter \csname mn@eprint@\@tempb\endcsname \expandafter{\@tempc}}}

\bibitem[\protect\citeauthoryear{Aasi et~al.,}{Aasi et~al.}{2015}]{aasi2015searches}
Aasi J.,  et~al., 2015, The Astrophysical Journal, 813, 39

\bibitem[\protect\citeauthoryear{Abadie et~al.,}{Abadie et~al.}{2012}]{PhysRevD.85.022001}
Abadie J.,  et~al., 2012, \mn@doi [Phys. Rev. D] {10.1103/PhysRevD.85.022001}, 85, 022001

\bibitem[\protect\citeauthoryear{Abbott et~al.,}{Abbott et~al.}{2016}]{Abbott_2016}
Abbott B.,  et~al., 2016, \mn@doi [Physical Review Letters] {10.1103/physrevlett.116.061102}, 116

\bibitem[\protect\citeauthoryear{Abbott et~al.,}{Abbott et~al.}{2017a}]{abbott2017first}
Abbott B.~P.,  et~al., 2017a, Physical Review D, 96, 122004

\bibitem[\protect\citeauthoryear{Abbott et~al.,}{Abbott et~al.}{2017b}]{PhysRevLett.119.161101}
Abbott B.~P.,  et~al., 2017b, \mn@doi [Phys. Rev. Lett.] {10.1103/PhysRevLett.119.161101}, 119, 161101

\bibitem[\protect\citeauthoryear{Abbott et~al.,}{Abbott et~al.}{2017c}]{Abbott_2017}
Abbott B.~P.,  et~al., 2017c, \mn@doi [The Astrophysical Journal] {10.3847/2041-8213/aa91c9}, 848, L12

\bibitem[\protect\citeauthoryear{Abbott et~al.,}{Abbott et~al.}{2019}]{abbott2019narrow}
Abbott B.~P.,  et~al., 2019, Physical Review D, 99, 122002

\bibitem[\protect\citeauthoryear{Abbott et~al.,}{Abbott et~al.}{2020}]{abbott2020gravitational}
Abbott R.,  et~al., 2020, The Astrophysical journal letters, 902, L21

\bibitem[\protect\citeauthoryear{Abbott et~al.,}{Abbott et~al.}{2021a}]{abbott2021all}
Abbott R.,  et~al., 2021a, Physical Review D, 104, 082004

\bibitem[\protect\citeauthoryear{Abbott et~al.,}{Abbott et~al.}{2021b}]{abbott2021diving}
Abbott R.,  et~al., 2021b, The Astrophysical Journal Letters, 913, L27

\bibitem[\protect\citeauthoryear{Abbott et~al.,}{Abbott et~al.}{2021c}]{abbott2021searches}
Abbott R.,  et~al., 2021c, The Astrophysical Journal, 921, 80

\bibitem[\protect\citeauthoryear{Abbott et~al.,}{Abbott et~al.}{2022a}]{abbott2022search}
Abbott R.,  et~al., 2022a, Physical Review D, 105, 022002

\bibitem[\protect\citeauthoryear{Abbott et~al.,}{Abbott et~al.}{2022b}]{PhysRevD.106.042003}
Abbott R.,  et~al., 2022b, \mn@doi [Phys. Rev. D] {10.1103/PhysRevD.106.042003}, 106, 042003

\bibitem[\protect\citeauthoryear{Abbott et~al.,}{Abbott et~al.}{2022c}]{PhysRevD.106.062002}
Abbott R.,  et~al., 2022c, \mn@doi [Phys. Rev. D] {10.1103/PhysRevD.106.062002}, 106, 062002

\bibitem[\protect\citeauthoryear{Abbott et~al.,}{Abbott et~al.}{2022d}]{abbott2022all}
Abbott R.,  et~al., 2022d, Physical Review D, 106, 102008

\bibitem[\protect\citeauthoryear{Abbott et~al.,}{Abbott et~al.}{2022e}]{Abbott_2022}
Abbott R.,  et~al., 2022e, \mn@doi [The Astrophysical Journal] {10.3847/1538-4357/ac6ad0}, 932, 133

\bibitem[\protect\citeauthoryear{Abbott et~al.,}{Abbott et~al.}{2022f}]{abbott2022narrowband}
Abbott R.,  et~al., 2022f, The Astrophysical Journal, 932, 133

\bibitem[\protect\citeauthoryear{Abbott et~al.,}{Abbott et~al.}{2022g}]{abbott2022searches}
Abbott R.,  et~al., 2022g, The Astrophysical Journal, 935, 1

\bibitem[\protect\citeauthoryear{Abbott et~al.}{Abbott et~al.}{2022h}]{LIGOScientific:2022enz}
Abbott R.,  et~al., 2022h, \mn@doi [Astrophys. J. Lett.] {10.3847/2041-8213/aca1b0}, 941, L30

\bibitem[\protect\citeauthoryear{{Andersson}}{{Andersson}}{1998}]{1998ApJ...502..708A}
{Andersson} N.,  1998, \mn@doi [\apj] {10.1086/305919}, \href {https://ui.adsabs.harvard.edu/abs/1998ApJ...502..708A} {502, 708}

\bibitem[\protect\citeauthoryear{Andersson, Kokkotas  \& Stergioulas}{Andersson et~al.}{1999}]{andersson1999relevance}
Andersson N.,  Kokkotas K.~D.,   Stergioulas N.,  1999, The Astrophysical Journal, 516, 307

\bibitem[\protect\citeauthoryear{Anzuini, Melatos, Dehman, Vigan{\`{o} }  \& Pons}{Anzuini et~al.}{2022}]{Anzuini_2022}
Anzuini F.,  Melatos A.,  Dehman C.,  Vigan{\`{o} } D.,   Pons J.~A.,  2022, \mn@doi [Monthly Notices of the Royal Astronomical Society] {10.1093/mnras/stac1353}, 515, 3014

\bibitem[\protect\citeauthoryear{Ashok et~al.,}{Ashok et~al.}{2021}]{Ashok_2021}
Ashok A.,  et~al., 2021, \mn@doi [The Astrophysical Journal] {10.3847/1538-4357/ac2582}, 923, 85

\bibitem[\protect\citeauthoryear{{Backus}}{{Backus}}{1986}]{1986RvGeo..24...75B}
{Backus} G.,  1986, \mn@doi [Reviews of Geophysics] {10.1029/RG024i001p00075}, \href {https://ui.adsabs.harvard.edu/abs/1986RvGeo..24...75B} {24, 75}

\bibitem[\protect\citeauthoryear{Bagchi, Ganguly, Layek, Sarkar  \& Srivastava}{Bagchi et~al.}{2023}]{bagchi2023qcd}
Bagchi P.,  Ganguly O.,  Layek B.,  Sarkar A.,   Srivastava A.~M.,  2023, QCD, Gravitational Waves, and Pulsars (\mn@eprint {arXiv} {2305.16850})

\bibitem[\protect\citeauthoryear{{Beznogov}, {Page}  \& {Ramirez-Ruiz}}{{Beznogov} et~al.}{2020}]{2020ApJ...888...97B}
{Beznogov} M.~V.,  {Page} D.,   {Ramirez-Ruiz} E.,  2020, \mn@doi [\apj] {10.3847/1538-4357/ab5fd6}, \href {https://ui.adsabs.harvard.edu/abs/2020ApJ...888...97B} {888, 97}

\bibitem[\protect\citeauthoryear{{Bildsten}}{{Bildsten}}{1998}]{1998ApJ...501L..89B}
{Bildsten} L.,  1998, \mn@doi [\apjl] {10.1086/311440}, \href {https://ui.adsabs.harvard.edu/abs/1998ApJ...501L..89B} {501, L89}

\bibitem[\protect\citeauthoryear{Bildsten, Cumming, Ushomirsky  \& Cutler}{Bildsten et~al.}{1997}]{bildsten1997oceanography}
Bildsten L.,  Cumming A.,  Ushomirsky G.,   Cutler C.,  1997, Oceanography of Accreting Neutron Stars: Non-Radial Oscillations and Periodic X-Ray Variability (\mn@eprint {arXiv} {astro-ph/9712358})

\bibitem[\protect\citeauthoryear{Brink \& Satchler}{Brink \& Satchler}{1962}]{brink1962angular}
Brink D.,  Satchler G.,  1962, Angular Momentum, by D.M. Brink and G.R. Satchler.
Oxford library of the physical sciences, Clarendon Press, \url {https://books.google.co.jp/books?id=i1TNswEACAAJ}

\bibitem[\protect\citeauthoryear{Carrascal, Estevez, Lee  \& Lorenzo}{Carrascal et~al.}{1991}]{Carrascal_1991}
Carrascal B.,  Estevez G.~A.,  Lee P.,   Lorenzo V.,  1991, \mn@doi [European Journal of Physics] {10.1088/0143-0807/12/4/007}, 12, 184

\bibitem[\protect\citeauthoryear{Chugunov \& Horowitz}{Chugunov \& Horowitz}{2010}]{10.1111/j.1745-3933.2010.00903.x}
Chugunov A.~I.,  Horowitz C.~J.,  2010, \mn@doi [Monthly Notices of the Royal Astronomical Society: Letters] {10.1111/j.1745-3933.2010.00903.x}, 407, L54

\bibitem[\protect\citeauthoryear{{Cie{\'s}lar}, {Bulik}, {Cury{\l}o}, {Sieniawska}, {Singh}  \& {Bejger}}{{Cie{\'s}lar} et~al.}{2021}]{2021A&A...649A..92C}
{Cie{\'s}lar} M.,  {Bulik} T.,  {Cury{\l}o} M.,  {Sieniawska} M.,  {Singh} N.,   {Bejger} M.,  2021, \mn@doi [\aap] {10.1051/0004-6361/202039503}, \href {https://ui.adsabs.harvard.edu/abs/2021A&A...649A..92C} {649, A92}

\bibitem[\protect\citeauthoryear{{Cook}, {Shapiro}  \& {Teukolsky}}{{Cook} et~al.}{1994}]{1994ApJ...424..823C}
{Cook} G.~B.,  {Shapiro} S.~L.,   {Teukolsky} S.~A.,  1994, \mn@doi [\apj] {10.1086/173934}, \href {https://ui.adsabs.harvard.edu/abs/1994ApJ...424..823C} {424, 823}

\bibitem[\protect\citeauthoryear{D'Onofrio et~al.,}{D'Onofrio et~al.}{2023}]{donofrio2023search}
D'Onofrio L.,  et~al., 2023, Search for gravitational wave signals from known pulsars in LIGO-Virgo O3 data using the 5n-vector ensemble method (\mn@eprint {arXiv} {2311.08229})

\bibitem[\protect\citeauthoryear{Dergachev \& Papa}{Dergachev \& Papa}{2021}]{Dergachev_2021}
Dergachev V.,  Papa M.~A.,  2021, \mn@doi [Physical Review D] {10.1103/physrevd.104.043003}, 104

\bibitem[\protect\citeauthoryear{{Duncan} \& {Thompson}}{{Duncan} \& {Thompson}}{1992}]{1992ApJ...392L...9D}
{Duncan} R.~C.,  {Thompson} C.,  1992, \mn@doi [\apjl] {10.1086/186413}, \href {https://ui.adsabs.harvard.edu/abs/1992ApJ...392L...9D} {392, L9}

\bibitem[\protect\citeauthoryear{Falxa et~al.,}{Falxa et~al.}{2023}]{falxa2023searching}
Falxa M.,  et~al., 2023, Monthly Notices of the Royal Astronomical Society, 521, 5077

\bibitem[\protect\citeauthoryear{Fantina, Ridder, Chamel  \& Gulminelli}{Fantina et~al.}{2020}]{Fantina_2020}
Fantina A.~F.,  Ridder S.~D.,  Chamel N.,   Gulminelli F.,  2020, \mn@doi [Astronomy \& Astrophysics] {10.1051/0004-6361/201936359}, 633, A149

\bibitem[\protect\citeauthoryear{{Friedman} \& {Schutz}}{{Friedman} \& {Schutz}}{1978}]{1978ApJ...221..937F}
{Friedman} J.~L.,  {Schutz} B.~F.,  1978, \mn@doi [\apj] {10.1086/156098}, \href {https://ui.adsabs.harvard.edu/abs/1978ApJ...221..937F} {221, 937}

\bibitem[\protect\citeauthoryear{{Fujisawa} \& {Kisaka}}{{Fujisawa} \& {Kisaka}}{2014}]{2014MNRAS.445.2777F}
{Fujisawa} K.,  {Kisaka} S.,  2014, \mn@doi [\mnras] {10.1093/mnras/stu1911}, \href {https://ui.adsabs.harvard.edu/abs/2014MNRAS.445.2777F} {445, 2777}

\bibitem[\protect\citeauthoryear{Gittins \& Andersson}{Gittins \& Andersson}{2019}]{Gittins_2019}
Gittins F.,  Andersson N.,  2019, \mn@doi [Monthly Notices of the Royal Astronomical Society] {10.1093/mnras/stz1719}, 488, 99

\bibitem[\protect\citeauthoryear{Gittins \& Andersson}{Gittins \& Andersson}{2021}]{Gittins_2021}
Gittins F.,  Andersson N.,  2021, \mn@doi [Monthly Notices of the Royal Astronomical Society] {10.1093/mnras/stab2048}, 507, 116

\bibitem[\protect\citeauthoryear{{Gittins}, {Andersson}  \& {Jones}}{{Gittins} et~al.}{2021}]{2021MNRAS.500.5570G}
{Gittins} F.,  {Andersson} N.,   {Jones} D.~I.,  2021, \mn@doi [\mnras] {10.1093/mnras/staa3635}, \href {https://ui.adsabs.harvard.edu/abs/2021MNRAS.500.5570G} {500, 5570}

\bibitem[\protect\citeauthoryear{{Goldreich} \& {Reisenegger}}{{Goldreich} \& {Reisenegger}}{1992}]{Goldreich_Reisenegger_1992}
{Goldreich} P.,  {Reisenegger} A.,  1992, \mn@doi [\apj] {10.1086/171646}, \href {http://adsabs.harvard.edu/abs/1992ApJ...395..250G} {395, 250}

\bibitem[\protect\citeauthoryear{Haskell, Jones  \& Andersson}{Haskell et~al.}{2006}]{Haskell_2006}
Haskell B.,  Jones D.~I.,   Andersson N.,  2006, \mn@doi [Monthly Notices of the Royal Astronomical Society] {10.1111/j.1365-2966.2006.10998.x}, 373, 1423

\bibitem[\protect\citeauthoryear{Haskell, Samuelsson, Glampedakis  \& Andersson}{Haskell et~al.}{2008}]{Haskell_2008}
Haskell B.,  Samuelsson L.,  Glampedakis K.,   Andersson N.,  2008, \mn@doi [Monthly Notices of the Royal Astronomical Society] {10.1111/j.1365-2966.2008.12861.x}, 385, 531

\bibitem[\protect\citeauthoryear{Haskell, Antonelli  \& Pizzochero}{Haskell et~al.}{2022}]{Haskell_2022}
Haskell B.,  Antonelli M.,   Pizzochero P.,  2022, \mn@doi [Universe] {10.3390/universe8120619}, 8, 619

\bibitem[\protect\citeauthoryear{Hessels, Ransom, Stairs, Freire, Kaspi  \& Camilo}{Hessels et~al.}{2006}]{doi:10.1126/science.1123430}
Hessels J. W.~T.,  Ransom S.~M.,  Stairs I.~H.,  Freire P. C.~C.,  Kaspi V.~M.,   Camilo F.,  2006, \mn@doi [Science] {10.1126/science.1123430}, 311, 1901

\bibitem[\protect\citeauthoryear{Horowitz \& Kadau}{Horowitz \& Kadau}{2009}]{Horowitz_2009}
Horowitz C.~J.,  Kadau K.,  2009, \mn@doi [Physical Review Letters] {10.1103/physrevlett.102.191102}, 102

\bibitem[\protect\citeauthoryear{Kalapotharakos, Wadiasingh, Harding  \& Kazanas}{Kalapotharakos et~al.}{2021}]{Kalapotharakos_2021}
Kalapotharakos C.,  Wadiasingh Z.,  Harding A.~K.,   Kazanas D.,  2021, \mn@doi [The Astrophysical Journal] {10.3847/1538-4357/abcec0}, 907, 63

\bibitem[\protect\citeauthoryear{Kaspi \& Beloborodov}{Kaspi \& Beloborodov}{2017}]{Kaspi_2017}
Kaspi V.~M.,  Beloborodov A.~M.,  2017, \mn@doi [Annual Review of Astronomy and Astrophysics] {10.1146/annurev-astro-081915-023329}, 55, 261

\bibitem[\protect\citeauthoryear{Kiendrebeogo et~al.,}{Kiendrebeogo et~al.}{2023}]{kiendrebeogo2023updated}
Kiendrebeogo R.~W.,  et~al., 2023, Updated observing scenarios and multi-messenger implications for the International Gravitational-wave Network's O4 and O5 (\mn@eprint {arXiv} {2306.09234})

\bibitem[\protect\citeauthoryear{{Kojima}}{{Kojima}}{1992}]{1992PhRvD..46.4289K}
{Kojima} Y.,  1992, \mn@doi [\prd] {10.1103/PhysRevD.46.4289}, \href {https://ui.adsabs.harvard.edu/abs/1992PhRvD..46.4289K} {46, 4289}

\bibitem[\protect\citeauthoryear{Kojima, Kisaka  \& Fujisawa}{Kojima et~al.}{2021}]{Kojima_2021}
Kojima Y.,  Kisaka S.,   Fujisawa K.,  2021, \mn@doi [Monthly Notices of the Royal Astronomical Society] {10.1093/mnras/stab1848}, 506, 3936

\bibitem[\protect\citeauthoryear{Kojima, Kisaka  \& Fujisawa}{Kojima et~al.}{2022}]{Kojima_2022}
Kojima Y.,  Kisaka S.,   Fujisawa K.,  2022, \mn@doi [Monthly Notices of the Royal Astronomical Society] {10.1093/mnras/stac036}, 511, 480

\bibitem[\protect\citeauthoryear{Kojima, Kisaka  \& Fujisawa}{Kojima et~al.}{2023}]{Kojima_2023}
Kojima Y.,  Kisaka S.,   Fujisawa K.,  2023, \mn@doi [The Astrophysical Journal] {10.3847/1538-4357/acc06b}, 946, 75

\bibitem[\protect\citeauthoryear{Lander}{Lander}{2021}]{Lander_2021}
Lander S.~K.,  2021, \mn@doi [Monthly Notices of the Royal Astronomical Society: Letters] {10.1093/mnrasl/slab086}, 507, L36

\bibitem[\protect\citeauthoryear{{Lattimer} \& {Prakash}}{{Lattimer} \& {Prakash}}{2007}]{2007PhR...442..109L}
{Lattimer} J.~M.,  {Prakash} M.,  2007, \mn@doi [\physrep] {10.1016/j.physrep.2007.02.003}, \href {https://ui.adsabs.harvard.edu/abs/2007PhR...442..109L} {442, 109}

\bibitem[\protect\citeauthoryear{Lindblom \& Owen}{Lindblom \& Owen}{2020}]{Lindblom_2020}
Lindblom L.,  Owen B.~J.,  2020, \mn@doi [Physical Review D] {10.1103/physrevd.101.083023}, 101

\bibitem[\protect\citeauthoryear{Lockitch \& Friedman}{Lockitch \& Friedman}{1999}]{Lockitch_1999}
Lockitch K.~H.,  Friedman J.~L.,  1999, \mn@doi [The Astrophysical Journal] {10.1086/307580}, 521, 764

\bibitem[\protect\citeauthoryear{Mastrano, Lasky  \& Melatos}{Mastrano et~al.}{2013}]{article}
Mastrano A.,  Lasky P.,   Melatos A.,  2013, \mn@doi [Monthly Notices of the Royal Astronomical Society] {10.1093/mnras/stt1131}, 434

\bibitem[\protect\citeauthoryear{Mastrano, Suvorov  \& Melatos}{Mastrano et~al.}{2015}]{Mastrano_2015}
Mastrano A.,  Suvorov A.~G.,   Melatos A.,  2015, \mn@doi [Monthly Notices of the Royal Astronomical Society] {10.1093/mnras/stu2671}, 447, 3475

\bibitem[\protect\citeauthoryear{Miller}{Miller}{2023}]{miller2023recent}
Miller A.~L.,  2023, Recent results from continuous gravitational wave searches using data from LIGO/Virgo/KAGRA's third observing run (\mn@eprint {arXiv} {2305.15185})

\bibitem[\protect\citeauthoryear{Morales \& Horowitz}{Morales \& Horowitz}{2022}]{Morales_2022}
Morales J.~A.,  Horowitz C.~J.,  2022, \mn@doi [Monthly Notices of the Royal Astronomical Society] {10.1093/mnras/stac3058}, 517, 5610

\bibitem[\protect\citeauthoryear{Owen}{Owen}{2005}]{PhysRevLett.95.211101}
Owen B.~J.,  2005, \mn@doi [Phys. Rev. Lett.] {10.1103/PhysRevLett.95.211101}, 95, 211101

\bibitem[\protect\citeauthoryear{Papa et~al.,}{Papa et~al.}{2020}]{Papa_2020}
Papa M.~A.,  et~al., 2020, \mn@doi [The Astrophysical Journal] {10.3847/1538-4357/ab92a6}, 897, 22

\bibitem[\protect\citeauthoryear{Papaloizou \& Pringle}{Papaloizou \& Pringle}{1978}]{10.1093/mnras/184.3.501}
Papaloizou J.,  Pringle J.~E.,  1978, \mn@doi [Monthly Notices of the Royal Astronomical Society] {10.1093/mnras/184.3.501}, 184, 501

\bibitem[\protect\citeauthoryear{Piccinni et~al.,}{Piccinni et~al.}{2020}]{Piccinni_2020}
Piccinni O.~J.,  et~al., 2020, \mn@doi [Physical Review D] {10.1103/physrevd.101.082004}, 101

\bibitem[\protect\citeauthoryear{{Roxburgh}}{{Roxburgh}}{1966}]{1966MNRAS.132..347R}
{Roxburgh} I.~W.,  1966, \mn@doi [\mnras] {10.1093/mnras/132.2.347}, \href {https://ui.adsabs.harvard.edu/abs/1966MNRAS.132..347R} {132, 347}

\bibitem[\protect\citeauthoryear{{Ruderman} \& {Sutherland}}{{Ruderman} \& {Sutherland}}{1973}]{1973NPhS..246...93R}
{Ruderman} M.~A.,  {Sutherland} P.~G.,  1973, \mn@doi [Nature Physical Science] {10.1038/physci246093a0}, \href {https://ui.adsabs.harvard.edu/abs/1973NPhS..246...93R} {246, 93}

\bibitem[\protect\citeauthoryear{Sieniawska \& Bejger}{Sieniawska \& Bejger}{2019}]{Sieniawska_2019}
Sieniawska M.,  Bejger M.,  2019, \mn@doi [Universe] {10.3390/universe5110217}, 5, 217

\bibitem[\protect\citeauthoryear{Smoluchowski \& Welch}{Smoluchowski \& Welch}{1970}]{PhysRevLett.24.1191}
Smoluchowski R.,  Welch D.~O.,  1970, \mn@doi [Phys. Rev. Lett.] {10.1103/PhysRevLett.24.1191}, 24, 1191

\bibitem[\protect\citeauthoryear{Stefanou, Pons  \& Cerd{\'{a} }-Dur{\'{a}}n}{Stefanou et~al.}{2022}]{Stefanou_2022}
Stefanou P.,  Pons J.~A.,   Cerd{\'{a} }-Dur{\'{a}}n P.,  2022, \mn@doi [Monthly Notices of the Royal Astronomical Society] {10.1093/mnras/stac3570}, 518, 6390

\bibitem[\protect\citeauthoryear{Steltner, Papa, Eggenstein, Prix, Bensch  \& Machenschalk}{Steltner et~al.}{2023}]{steltner2023deep}
Steltner B.,  Papa M.~A.,  Eggenstein H.~B.,  Prix R.,  Bensch M.,   Machenschalk B.,  2023, Deep Einstein@Home all-sky search for continuous gravitational waves in LIGO O3 public data (\mn@eprint {arXiv} {2303.04109})

\bibitem[\protect\citeauthoryear{Sur \& Haskell}{Sur \& Haskell}{2021}]{Sur_2021}
Sur A.,  Haskell B.,  2021, \mn@doi [Monthly Notices of the Royal Astronomical Society] {10.1093/mnras/stab307}, 502, 4680

\bibitem[\protect\citeauthoryear{{Suwa}}{{Suwa}}{2014}]{2014PASJ...66L...1S}
{Suwa} Y.,  2014, \mn@doi [\pasj] {10.1093/pasj/pst030}, \href {https://ui.adsabs.harvard.edu/abs/2014PASJ...66L...1S} {66, L1}

\bibitem[\protect\citeauthoryear{Ushomirsky}{Ushomirsky}{2000}]{Ushomirsky_2000}
Ushomirsky G.,  2000, in {AIP} Conference Proceedings. {AIP}, \mn@doi{10.1063/1.1291841}, \url {https://doi.org/10.1063\%2F1.1291841}

\bibitem[\protect\citeauthoryear{{Ushomirsky}, {Cutler}  \& {Bildsten}}{{Ushomirsky} et~al.}{2000}]{Ushomirsky_et_al_2000}
{Ushomirsky} G.,  {Cutler} C.,   {Bildsten} L.,  2000, \mn@doi [\mnras] {10.1046/j.1365-8711.2000.03938.x}, \href {https://ui.adsabs.harvard.edu/abs/2000MNRAS.319..902U} {319, 902}

\bibitem[\protect\citeauthoryear{Varma, Müller  \& Schneider}{Varma et~al.}{2022}]{Varma_2022}
Varma V.,  Müller B.,   Schneider F. R.~N.,  2022, \mn@doi [Monthly Notices of the Royal Astronomical Society] {10.1093/mnras/stac3247}, 518, 3622

\bibitem[\protect\citeauthoryear{Viganò, Rea, Pons, Perna, Aguilera  \& Miralles}{Viganò et~al.}{2013}]{10.1093/mnras/stt1008}
Viganò D.,  Rea N.,  Pons J.~A.,  Perna R.,  Aguilera D.~N.,   Miralles J.~A.,  2013, \mn@doi [Monthly Notices of the Royal Astronomical Society] {10.1093/mnras/stt1008}, 434, 123

\bibitem[\protect\citeauthoryear{Viganò, Pons, Miralles  \& Rea}{Viganò et~al.}{2015}]{viganò2015magnetic}
Viganò D.,  Pons J.~A.,  Miralles J.~A.,   Rea N.,  2015, Magnetic fields in Neutron Stars (\mn@eprint {arXiv} {1501.06735})

\bibitem[\protect\citeauthoryear{{Wagoner}}{{Wagoner}}{1984}]{1984ApJ...278..345W}
{Wagoner} R.~V.,  1984, \mn@doi [\apj] {10.1086/161798}, \href {https://ui.adsabs.harvard.edu/abs/1984ApJ...278..345W} {278, 345}

\bibitem[\protect\citeauthoryear{Wette}{Wette}{2023}]{wette2023searches}
Wette K.,  2023, Searches for continuous gravitational waves from neutron stars: A twenty-year retrospective (\mn@eprint {arXiv} {2305.07106})

\bibitem[\protect\citeauthoryear{White, Burrows, Coleman  \& Vartanyan}{White et~al.}{2022}]{White_2022}
White C.~J.,  Burrows A.,  Coleman M. S.~B.,   Vartanyan D.,  2022, \mn@doi [The Astrophysical Journal] {10.3847/1538-4357/ac4507}, 926, 111

\bibitem[\protect\citeauthoryear{Wynn \& Dong}{Wynn \& Dong}{1999}]{Wynn_1999}
Wynn C. G.~H.,  Dong L.,  1999, \mn@doi [Monthly Notices of the Royal Astronomical Society] {10.1046/j.1365-8711.1999.02703.x}, 308, 153

\bibitem[\protect\citeauthoryear{Yadav, Mishra, Sarkar  \& Singh}{Yadav et~al.}{2022}]{yadav2022thermal}
Yadav S.,  Mishra M.,  Sarkar T.~G.,   Singh C.~R.,  2022, Thermal Evolution and Emission Properties of Strongly Magnetized Neutron Star (\mn@eprint {arXiv} {2212.11652})

\bibitem[\protect\citeauthoryear{Zhu et~al.,}{Zhu et~al.}{2016}]{Zhu_2016}
Zhu S.~J.,  et~al., 2016, \mn@doi [Physical Review D] {10.1103/physrevd.94.082008}, 94

\makeatother
\end{thebibliography}




\appendix

\onecolumn

\section{Orientation of the dipolar magnetic field}\label{sb:4.1}
There is an issue with considering a dipolar ($(l,m) = (1,0)$) magnetic field inducing quadrupolar ($(l,m) = (2,2)$) deformations which are responsible for the emission of CGWs. We have demonstrated how the $l=1$ multipole of the magnetic field can couple to produce the $l=2$ multipolar deformations in Appendix \ref{appendix A}. The issue lies with the $m$-harmonics, in which we are interested in the $m=0$ harmonic of the magnetic field inducing $m=2$ harmonic of the neutron star mountain. One of the possibilities in which such an $m$-harmonic coupling can be observed is when the axis of the dipolar magnetic field is misaligned with the assumed spin axis of the neutron star. The logic behind this assumption is the fact that CGWs will be emitted only when the neutron star deformed away from axisymmetry starts rotating at a particular angular frequency. The reason why we are not considering a dynamically rotating star in this scenario is that we can ignore $(l,m) = (2,0)$ harmonic centrifugal deformations generated by the rotation of the star. Thus, we can always assume that the spin axis is along the $\hat{\phi}$ direction of the spherical harmonics coordinate system that we are using. The misalignment (or the orientation) of the axis of the dipolar magnetic field of our star is thus considered relative to this assumed spin axis of the neutron star. 

Borrowing the framework from \cite{Wynn_1999}, we describe the orientation of the axis of the dipolar magnetic field with respect to the spin axis using the Euler angles $(\alpha,\beta, \gamma = 0)$. We assume the coordinate frame stuck to the spin axis is labelled with the angular coordinates $\theta,\phi$ whereas the coordinate frame stuck to the magnetic field's axis is labelled with the angular coordinates $\theta',\phi'$. The spherical harmonics transform from the axis of the dipolar magnetic field to the spin axis as \begin{equation}
    Y_{lm}(\theta,\phi) = \sum_{m'}\mathcal{D}_{m'm}^{(l)}(\alpha,\beta)Y_{lm'}(\theta', \phi')
\end{equation} where the Wigner-$\mathcal{D}$ function is defined as \begin{equation}\mathcal{D}_{m'm}^{(l)}(\alpha,\beta) = e^{im'\alpha}[(l+m)!(l-m)!(l+m')!(l-m')!]^{1/2}\sum_{p}\frac{(-1)^{l+m'-p}(\cos{\beta/2})^{2p-m-m'}(\sin{\beta/2})^{2l-2p+m+m'}}{p!(l+m-p)!(l+m'-p)!(p-m-m')!},
\end{equation} and in our case $l=2$, $m=2$, $m'=0$, $\alpha$ is the angle between $\phi$ and $\phi'$ and $\beta$ is the angle between $\theta$ and $\theta'$. The angle $\alpha$ is arbitrary and $\beta$ is set to $\beta = \pi/2$ to maximize the amplitude of the CGW emissions from the neutron star (\citealt{Sur_2021}). For the Wigner-$\mathcal{D}$ function, $p$ is an integer and lies in the range of $p \in [\max(0,m'-m),\min(l-m,l+m')]$, which for our case is $p = 0$ which gives us the expression \begin{equation}
    \mathcal{D}^{(2)}_{0,2}(\alpha,\pi/2) = \frac{1}{16\sqrt{6}}. 
\end{equation} We ignore this multiplicative factor in our perturbation equations in later sections for the sake of simplicity.

\section{Vector Spherical Harmonics Integrals}\label{appendix A}

While calculating the Lorentz force in section \ref{section:5} and section \ref{section:6}, we encounter multiple triple integrals of vector spherical harmonics. These integrals over three spherical harmonics (and/or their derivatives) are obtained when we use the orthogonality condition of the vector spherical harmonics to obtain perturbation equations which are independent of the angular coordinates ($\theta, \phi$). These are easily simplified using the contraction rules of spherical harmonics (\citealt{brink1962angular}) and their algebraic relations as used by \cite{Lockitch_1999}. This gives us relations for the coupling of $l$-harmonics with their adjacent harmonics. We remove the summation over spherical harmonics using the orthogonality condition obtained from these integral relations. On the other hand, the summation due to the contraction rule is removed since we choose only the dipolar ($(l,m) = (1,0)$) magnetic fields and quadrupolar ($(l,m) = (2,2)$) displacement vectors and Lorentz force while ignoring the effect of other higher-order multipoles completely. This is a choice made for simplification and can be discarded in favour of more general formulations in other works. The triple integrals of spherical harmonics and their associated matrices expressing the coupling of different l-harmonics are given here using the Wigner 3j-Symbol as:

\begin{equation}
    \mathbf{A}_{1} \equiv \int Y^{*}_{l_{3}m_{3}}Y_{l_{1}m_{1}}Y_{l_{2}m_{2}}d\Omega = \sqrt{\frac{(2l_{1}+1)(2l_{2}+1)(2l_{3}+1)}{4\pi}}\begin{pmatrix} l_{1} & l_{2} & l_{3} \\ 0 & 0 & 0 \end{pmatrix}\begin{pmatrix} l_{1} & l_{2} & l_{3} \\ m_{1} & m_{2} & -m_{3} \end{pmatrix}
\end{equation} \[
    \mathbf{A}_{2} \equiv \int Y^{*}_{l_{3}m_{3}} (\Vec{\nabla}Y_{l_{1}m_{1}}\cdot\Vec{\nabla}Y_{l_{2}m_{2}})d\Omega =\frac{1}{r^{2}}\Bigg[-m_{1}m_{2}\sqrt{\frac{(2l_{1}+1)(2l_{2}+1)(2l_{3}+1)}{4\pi}}\begin{pmatrix} l_{1} & l_{2} & l_{3} \\ 0 & 0 & 0 \end{pmatrix}\begin{pmatrix} l_{1} & l_{2} & l_{3} \\ m_{1} & m_{2} & -m_{3} \end{pmatrix}\]\[ + l_{1}l_{2}Q_{l_{1}+1}Q_{l_{2}+1}\sqrt{\frac{(2l_{3}+1)(2l_{1}+3)(2l_{2}+3)}{4\pi}}\begin{pmatrix} l_{1}+1 & l_{2}+1 & l_{3} \\ 0 & 0 & 0 \end{pmatrix}\begin{pmatrix} l_{1}+1 & l_{2}+1 & l_{3} \\ m_{1} & m_{2} & -m_{3} \end{pmatrix} +\]\[ (l_{1}+1)(l_{2}+1)Q_{l_{1}}Q_{l_{2}}\sqrt{\frac{(2l_{3}+1)(2l_{1}-1)(2l_{2}-1)}{4\pi}}\begin{pmatrix} l_{1}-1 & l_{2}-1 & l_{3} \\ 0 & 0 & 0 \end{pmatrix}\begin{pmatrix} l_{1}-1 & l_{2}-1 & l_{3} \\ m_{1} & m_{2} & -m_{3} \end{pmatrix} -\]\begin{equation} 2l_{1}(l_{2}+1)Q_{l_{1}+1}Q_{l_{2}}\sqrt{\frac{(2l_{3}+1)(2l_{2}-1)(2l_{1}+3)}{4\pi}}\begin{pmatrix} l_{1}+1 & l_{2}-1 & l_{3} \\ 0 & 0 & 0 \end{pmatrix}\begin{pmatrix} l_{1}+1 & l_{2}-1 & l_{3} \\ m_{1} & m_{2} & -m_{3} \end{pmatrix}\Bigg]
\end{equation} \[
    \mathbf{B}_{1} \equiv \int \Vec{\nabla}Y^{*}_{l_{3}m_{3}}\cdot Y_{l_{1}m_{1}}\Vec{\nabla}Y_{l_{2}m_{2}}d\Omega = \frac{1}{r}\Bigg[m_{2}m_{3}\sqrt{\frac{(2l_{1}+1)(2l_{2}+1)(2l_{3}+1)}{4\pi}}\begin{pmatrix} l_{1} & l_{2} & l_{3} \\ 0 & 0 & 0 \end{pmatrix}\begin{pmatrix} l_{1} & l_{2} & l_{3} \\ m_{1} & m_{2} & -m_{3} \end{pmatrix}\]\[ + l_{1}l_{3}Q_{l_{1}+1}Q_{l_{3}+1}\sqrt{\frac{(2l_{2}+1)(2l_{1}+3)(2l_{2}+3)}{4\pi}}\begin{pmatrix} l_{1}+1 & l_{2} & l_{3}+1 \\ 0 & 0 & 0 \end{pmatrix}\begin{pmatrix} l_{1}+1 & l_{2} & l_{3}+1 \\ m_{1} & m_{2} & -m_{3} \end{pmatrix} +\]\[ (l_{1}+1)(l_{3}+1)Q_{l_{1}}Q_{l_{2}}\sqrt{\frac{(2l_{2}+1)(2l_{1}-1)(2l_{3}-1)}{4\pi}}\begin{pmatrix} l_{1}-1 & l_{2} & l_{3}-1 \\ 0 & 0 & 0 \end{pmatrix}\begin{pmatrix} l_{1}-1 & l_{2} & l_{3}-1 \\ m_{1} & m_{2} & -m_{3} \end{pmatrix} -\]\[ l_{1}(l_{3}+1)Q_{l_{1}+1}Q_{l_{3}}\sqrt{\frac{(2l_{2}+1)(2l_{3}-1)(2l_{1}+3)}{4\pi}}\begin{pmatrix} l_{1}+1 & l_{2} & l_{3}-1 \\ 0 & 0 & 0 \end{pmatrix}\begin{pmatrix} l_{1}+1 & l_{2} & l_{3}-1 \\ m_{1} & m_{2} & -m_{3} \end{pmatrix} - \]\begin{equation}l_{3}(l_{1}+1)Q_{l_{3}+1}Q_{l_{1}}\sqrt{\frac{(2l_{2}+1)(2l_{1}-1)(2l_{3}+3)}{4\pi}}\begin{pmatrix} l_{1}-1 & l_{2} & l_{3}+1 \\ 0 & 0 & 0 \end{pmatrix}\begin{pmatrix} l_{1}-1 & l_{2} & l_{3}+1 \\ m_{1} & m_{2} & -m_{3} \end{pmatrix}\Bigg]
\end{equation} \begin{equation}
\mathbf{C}_{1} \equiv \int(\hat{r}\cross\Vec{\nabla}Y^{*}_{l_{3}m_{3}})\cdot(Y_{l_{1}m_{1}}\hat{r}\cross\Vec{\nabla}Y_{l_{2}m_{2}})d\Omega = \mathbf{B}_{1}
\end{equation}
where $Q_{l}$ is shorthand for $Q_{lm}$ (the $m$-harmonic is irrelevant here as it will be set to zero for all the integrals as well as showing no coupling as a result of these integrals), which is defined in \cite{Lockitch_1999} as \begin{equation}
    Q_{lm} = \Bigg[\frac{(l+m)(l-m)}{(2l-1)(2l+1)}\Bigg]^{1/2} = \Bigg[\frac{l^{2}}{4l^{2}-1}\Bigg]^{1/2}.
\end{equation}

\section{Equivalence of Multipolar decomposition and Poloidal-Toroidal decomposition}\label{appendix B}
The Poloidal and Toroidal components of a solenoidal vector field like the magnetic field can be expressed as \cite{1986RvGeo..24...75B}\begin{equation}\label{eq:a1}
    \Vec{B} = \Vec{\nabla}\cross(\hat{r}i\Psi) + \Vec{\nabla}\cross(\Vec{\nabla}\cross(\hat{r}\Phi))
\end{equation} where $\Psi(r,\theta,\phi)$ and $\Phi(r,\theta,\phi)$ are scalar fields pointed in the radially outward directions. We decompose them using the spherical harmonics basis as \begin{equation}
     \Psi = \Psi_{lm}Y_{lm}, \Phi = \Phi_{lm}Y_{lm}
\end{equation} since they are scalar fields on a spherical background. The $i$ symbol denotes that the harmonics corresponding to each scalar field need not be the same, as will also be apparent later. It also leads to the separation of the poloidal and toroidal components of the magnetic field more naturally. We now insert the above decompositions into \ref{eq:a1} to obtain \begin{equation}
    \Vec{B} = -i\Psi_{lm}(\hat{r}\cross\Vec{\nabla}Y_{lm}) + \beta^{2}\Phi_{lm}Y_{lm}\hat{r} + \frac{d\Phi}{dr}\Vec{\nabla}Y_{lm}.
\end{equation} We can now re-label the radial functions in the above expression as $\Psi_{lm} = -B_{\cross}\frac{r}{\beta}$, $\beta^{2}\Phi_{lm} = B_{r}$ and $\frac{d\Phi}{dr} = B_{\perp}\frac{r}{\beta}$ to obtain our multipolar decomposition of the magnetic field as expressed in equation \ref{eq:4}. We notice that there is a relation between the radial functions $B_{r}$ and $B_{\perp}$. This relation is a consequence of the magnetic field being solenoidal, i.e. it is a consequence of the divergence-free condition of the magnetic field ($\Vec{\nabla}\cdot\Vec{B} = 0$), which relates the radial and angular components of the magnetic field.

\section{Magnetoelastic-dynamical equilibrium in the crust of the neutron star}\label{MED appendix}
One of the cornerstones of calculating the spatial configuration of the magnetic field in a neutron star is the MHD equilibrium condition \ref{eq:MHD}. This is always true for a fluid star since the curl of every other term in the perturbed Euler equation vanishes. However, this is not true for the crust once it has solidified, strained and deformed by the magnetic field. At this stage, it is essential to consider the Magnetoelastic-dynamical (MED) equilibrium condition since the curl of the force generated by a strained elastic crust does not vanish. This is given as \cite{Kojima_2021}\begin{equation}\Vec{\nabla}\cross\Bigg(\frac{(\Vec{\nabla}\cross\Vec{B})\cross\Vec{B}}{4\pi\rho} + \Vec{\nabla}\cdot(\mathbf{t})\Bigg) = 0.
\end{equation} When we consider the MED equilibrium condition instead of the MHD equilibrium condition, equation \ref{eq:MHDfull1} will be modified as follows: \[\Bigg(B_{r} +  r\frac{dB_{r}}{dr} - \frac{rB_{r}}{\rho}\frac{d\rho}{dr} +  \frac{r^{2}B_{\perp}}{\beta}\Bigg)\Bigg(\frac{dB_{\perp}}{dr} + \frac{B_{\perp}}{r} -   \frac{B_{r}}{r}\Bigg) + rB_{r}\Bigg(\frac{d^{2}B_{\perp}}{dr^{2}} + \frac{1}{r}\frac{dB_{\perp}}{dr} - \frac{B_{\perp}}{r^{2}} + \frac{B_{r}}{r^{2}} - \frac{1}{r}\frac{dB_{r}}{dr}\Bigg) + \]\begin{equation} \frac{1}{\rho}\frac{d\mu}{dr}\Bigg(-\frac{1}{3r}\frac{d\xi_{r}}{dr} - \frac{\xi_{r}}{3r^{2}} - \frac{2\beta\xi_{\perp}}{3r^{2}} + \frac{d^{2}\xi_{\perp}}{dr^{2}} + \frac{2}{r}\frac{d\xi_{\perp}}{dr}\Bigg) + \frac{\mu}{\rho}\Bigg(-\frac{1}{r}\frac{d^{2}\xi_{r}}{dr^{2}} + \frac{\beta^{2}\xi_{r}}{r^{3}} + \frac{3}{r}\frac{d^{2}\xi_{\perp}}{dr^{2}} + \frac{d^{3}\xi_{\perp}}{dr^{3}} - \frac{5\beta^{2}\xi_{\perp}}{r^{3}} + \frac{\beta^{2}}{r^{2}}\frac{d\xi_{\perp}}{dr}\Bigg) = 0\end{equation} for a poloidal magnetic field and for a toroidal magnetic field, equation \ref{eq:MHDfull2} will be modified as \[ \frac{B^{2}_{\cross}}{\rho}\frac{d\rho}{dr} - 3B_{\cross}\frac{dB_{\cross}}{dr} - \frac{2B^{2}_{\cross}}{r} +  \frac{1}{\rho}\frac{d\mu}{dr}\Bigg(-\frac{1}{3r}\frac{d\xi_{r}}{dr} - \frac{\xi_{r}}{3r^{2}} - \frac{2\beta\xi_{\perp}}{3r^{2}} + \frac{d^{2}\xi_{\perp}}{dr^{2}} + \frac{2}{r}\frac{d\xi_{\perp}}{dr}\Bigg) + \]\begin{equation} \frac{\mu}{\rho}\Bigg(-\frac{1}{r}\frac{d^{2}\xi_{r}}{dr^{2}} + \frac{\beta^{2}\xi_{r}}{r^{3}} + \frac{3}{r}\frac{d^{2}\xi_{\perp}}{dr^{2}} + \frac{d^{3}\xi_{\perp}}{dr^{3}} - \frac{5\beta^{2}\xi_{\perp}}{r^{3}} + \frac{\beta^{2}}{r^{2}}\frac{d\xi_{\perp}}{dr}\Bigg) = 0.\end{equation}
The new extra terms included in this equation, most of which are operations on $\xi_{r}$ and $\xi_{\perp}$, lead to the coupling of the equations governing the magnetic field's geometrical configuration and the crustal deformations. The only way to deal with this completely is to solve the MED equilibrium condition above, the divergence-free condition of the magnetic field, and the perturbation equations governing the crustal deformations together. This is a challenging task numerically and is hence left to be done in the near future. 
In this paper, we have stuck to the MHD equilibrium condition instead of utilizing the MED equilibrium condition for three reasons: \begin{enumerate}
    \item In the force-based approach of mountain formation, we justify the use of MHD equilibrium instead of MED equilibrium since the extra terms introduced by assuming MED equilibrium are all much smaller compared to the rest and are safely ignored. As demonstrated in figure \ref{fig3}, the fluid neutron star in state \textbf{S} is deformed by the magnetic field to a state \textbf{SM}, which is in MHD equilibrium. The star state \textbf{SC} with a solidified crust is deformed by the magnetic field to a deformed star state \textbf{SCM}, which is in MED equilibrium. When we subtract state \textbf{SM} from state \textbf{SCM}, we obtain the perturbative equations of crustal deformations that we are interested in solving. Since the equations governing the spatial configuration of the magnetic fields are also coupled with these perturbative equations, we must also consider them together in this procedure which leads to the emergence of a third-order perturbative equation (comprising of all the aforementioned extra terms) which can be safely ignored for our calculations. This is because this differential equation is proportional to $\frac{\mu}{\rho}\xi = k\xi$ (\ref{eq:kappa}), which is much smaller compared to the rest of the perturbation equations by multiple orders of magnitude. This subtraction of states also leads to the decoupling of the equations governing the spatial configuration of the magnetic fields and the perturbation equations describing the crustal deformations in our star. 
    \begin{figure}
        \centering
        \includegraphics[width=11cm]{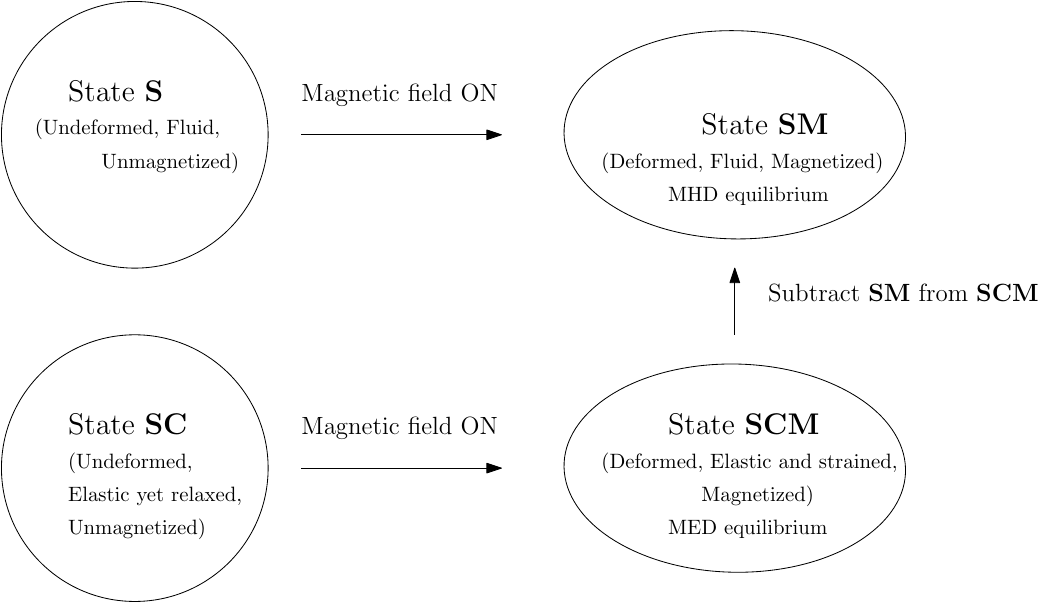}
        \caption{Formation of a magnetic mountain using the Gittins scheme assuming MED equilibrium}
        \label{fig3}
    \end{figure}
    \item In the formation history-based approach of mountain formation, we know that the strain of the elastic crust emerges as the force deforming the crust dissipates to zero. Since we are only interested in the final deformed state \textbf{SE} of the star, we do not need to consider the MED equilibrium at any stage of the process. When the magnetic field deforms the star initially, it is in MHD equilibrium with the fluid star in state \textbf{SM} but once the elastic crust of the star solidifies in state \textbf{SME}, the strain in it is still zero and hence considering MED equilibrium is irrelevant for our intents and purposes. We consider non-zero strain only at the state \textbf{SE}, where the magnetic field itself has dissipated away and consideration of MED equilibrium is no longer necessary. 
    \item There is however a problem when we start considering states between star state \textbf{SME} and star state \textbf{SE}. These intermittent states are irrelevant while considering the perturbation equations governing the crustal deformations but their effect on the spatial configuration of magnetic fields must be considered carefully. In these intermittent states, the magnetic field has decayed away partially and hence, non-zero strain has also built up in the elastic crust of the star. The safest way to explain these states and the configuration of the magnetic fields in these states is to use MED equilibrium. Fortunately, we do not have to deal with these states at all and hence, we can assume that for our intents and purposes, the magnetic field, once in MHD equilibrium, is now insusceptible to any other changes in the fluid configuration. 
\end{enumerate}

\bsp	
\label{lastpage}
\end{document}